\documentclass[a4paper,fleqn,usenatbib]{mnras}

\usepackage[T1]{fontenc}
\usepackage{ae,aecompl}

\usepackage{graphicx}	
\usepackage{amsmath}	
\usepackage{amssymb}	
\usepackage{multirow}
\usepackage{booktabs}
\usepackage{epsfig}
\usepackage{longtable}
\usepackage{supertabular}
\usepackage{caption}
\usepackage{ulem}

\def\arcsec{\hbox{$^{\prime\prime}$}}
\def\lesssim{\mathrel{\hbox{\rlap{\hbox{\lower3pt\hbox{$\sim$}}}\hbox{\raise1pt\hbox{$<$}}}}}
\def\gtrsim{\mathrel{\hbox{\rlap{\hbox{\lower3pt\hbox{$\sim$}}}\hbox{\raise1pt\hbox{$>$}}}}}
\def\wifes{WiFeS}

\newcommand{\msun}{M$_\odot$}
\newcommand{\rsun}{R$_\odot$}
\def\co{$^{56}{\rm Co}$}
\def\ni{$^{56}{\rm Ni}$}
\newcommand{\ha}{H$\alpha$}

\newcommand{\hi}{H\,{\footnotesize I}}
\newcommand{\hei}{He\,{\footnotesize I}}
\newcommand{\hh}{H\,{\footnotesize II}}
\newcommand{\oi}{[O\,{\footnotesize I}]}
\newcommand{\caii}{Ca\,{\footnotesize II}}
\newcommand{\feii}{Fe\,{\footnotesize II}}
\newcommand{\niii}{Ni\,{\footnotesize II}}
\newcommand{\mgi}{Mg\,{\footnotesize I}}
\newcommand{\sii}{Si\,{\footnotesize I}}
\newcommand{\siii}{Si\,{\footnotesize II}}
\newcommand{\nai}{Na\,{\footnotesize I}}
\newcommand{\naid}{Na\,{\footnotesize I\,D}}

\title[SN 2013ej]{450 Days of Type II SN 2013ej in Optical and Near-Infrared}

\author[F.~Yuan et al]
{\parbox{\textwidth}{
Fang~Yuan,$^{1,2}$\thanks{E-mail:
fang.yuan@anu.edu.au}
A.~Jerkstrand$^{3}$,
S.~Valenti$^{4}$,
J.~Sollerman$^{5}$,
I.~R.~Seitenzahl$^{1,2}$,
A.~Pastorello$^{6}$,
S.~Schulze$^{7,8}$,
T.-W.~Chen$^{9}$,
M.~J.~Childress$^{1,2,10}$,
M.~Fraser$^{11}$,
C.~Fremling$^{5}$,
R.~Kotak$^{3}$,
A.~J.~Ruiter$^{1,2}$,
B.~P.~Schmidt$^{1,2}$,
S.~J.~Smartt$^{3}$,
F.~Taddia$^{5}$,
G.~Terreran$^{3,6}$,
B.~E.~Tucker$^{1,2}$,
C.~Barbarino$^{12,6}$,
S.~Benetti$^{6}$,
N.~Elias-Rosa$^{6}$,
A.~Gal-Yam$^{13}$,
D.~A.~Howell$^{14,15}$,
C.~Inserra$^{3}$,
E.~Kankare$^{3}$,
M.~Y.~Lee$^{16}$,
K.~L.~Li$^{17,18}$,
K.~Maguire$^{3}$,
S.~Margheim$^{19}$,
A.~Mehner$^{20}$,
P.~Ochner$^{6}$,
M.~Sullivan$^{10}$,
L.~Tomasella$^{6}$,
D.~R.~Young$^{3}$,
}\vspace{0.4cm}\\
$^{1}$Research School of Astronomy and Astrophysics, Australian National University, Canberra, ACT 2611, Australia\\
$^{2}$ARC Centre of Excellence for All-sky Astrophysics (CAASTRO) \\
$^{3}$Astrophysics Research Centre, School of Mathematics and Physics, Queen's University Belfast, Belfast BT7 1NN, UK \\
$^{4}$Department of Physics, University of California, Davis, CA 95616, USA\\
$^{5}$The Oskar Klein Centre, Department of Astronomy, Stockholm University, AlbaNova, 10691, Stockholm, Sweden\\
$^{6}$INAF - Osservatorio Astronomico di Padova, Vicolo dellÕOsservatorio 5, 35122 Padova, Italy\\
$^{7}$Instituto de Astrof\'isica, Facultad de F\'isica, Pontificia Universidad Cat\'olica de Chile, Vicu\~{n}a Mackenna 4860, 7820436 Macul, Santiago, Chile\\
$^{8}$Millennium Institute of Astrophysics, Vicu\~{n}a Mackenna 4860, 7820436 Macul, Santiago, Chile\\
$^{9}$Max-Planck-Institut f{\"u}r Extraterrestrische Physik, Giessenbachstra\ss e 1, 85748, Garching, Germany\\
$^{10}$Department of Physics and Astronomy, University of Southampton, Southampton, SO17 1BJ, UK\\
$^{11}$Institute of Astronomy, University of Cambridge, Madingley Rd., Cambridge, CB3 0HA, UK\\
$^{12}$Dip. di Fisica and ICRA, Sapienza Universita' di Roma, Piazzale Aldo Moro 5, I-00185 Rome, Italy\\
$^{13}$Benoziyo Center for Astrophysics, Weizmann Institute of Science, 76100 Rehovot, Israel\\
$^{14}$Las Cumbres Observatory Global Telescope Network, 6740 Cortona Dr., Suite 102, Goleta, CA 93117, USA\\
$^{15}$Department of Physics, University of California, Santa Barbara, Broida Hall, Mail Code 9530, Santa Barbara, CA 93106-9530, USA\\
$^{16}$Taipei First Girls' High School, Taiwan\\
$^{17}$Department of Physics and Astronomy, Michigan State University, East Lansing, MI48824, USA\\
$^{18}$Institute of Astronomy and Department of Physics, National Tsing Hua University, Hsinchu, Taiwan\\
$^{19}$Gemini Observatory, Southern Operations Center, Casilla 603, La Serena, Chile\\
$^{20}$European Southern Observatory, Alonso de C\'{o}rdova 3107, Vitacura Casilla 19001, Santiago, Chile\\
}
\date{Accepted XXX. Received YYY; in original form ZZZ}

\pubyear{2016}

\begin{document}
\label{firstpage}
\pagerange{\pageref{firstpage}--\pageref{lastpage}}
\maketitle

\begin{abstract}
We present optical and near-infrared photometric and spectroscopic observations of SN 2013ej, in galaxy M74, from 1 to 450 days after the explosion.
SN 2013ej is a hydrogen-rich supernova, classified as a Type IIL due to its relatively fast decline following the initial peak.
It has a relatively high peak luminosity (absolute magnitude M$_V$ = -17.6) but a small {\ni} production of $\sim$0.023 {\msun}.
Its photospheric evolution is similar to other Type II SNe, with shallow absorption in the $H_{\alpha}$ profile typical for a Type IIL.
During transition to the radioactive decay tail at $\sim$100 days, we find the SN to grow bluer in $B-V$ colour, in contrast to some other Type II supernovae.
At late times, the bolometric light curve declined faster than expected from {\co} decay and we observed unusually broad and asymmetric nebular emission lines.
Based on comparison of nebular emission lines most sensitive to the progenitor core mass, we find our observations are 
best matched to synthesized spectral models with a M$_\mathrm{ZAMS} = 12 -15$ {\msun} progenitor.
The derived mass range is similar to but not higher than the mass estimated for Type IIP progenitors.
This is against the idea that Type IIL are from more massive stars.
Observations are consistent with the SN having a progenitor with a relatively low-mass envelope.

\end{abstract}




\section{Introduction}

Massive stars die young and spectacularly as core-collapse supernovae (CCSNe). 
Luminous SN explosions allow us to study individual stars in distant galaxies, unraveling their life stories and those of their host galaxies. 
How a star dies depends on how it was born in its specific host environment. 
The energy and heavy elements released during the SN explosion in turn enrich the environment and affect the next generation of stars. 
A central piece to understand this ecology is the mapping between types of progenitor stars and the observed variety of SNe.

About half of all CCSNe show hydrogen P-Cygni profiles in their spectra and display a prolonged plateau phase in their optical light curve \citep{2011MNRAS.412.1441L,2011MNRAS.412.1522S}. 
These SNe are accordingly designated as Type II-Plateau (IIP). 
Type IIPs are believed to be produced by relatively low-mass progenitors that have retained their hydrogen envelope until the explosion. 
The most concrete evidence for this connection is the direct detection of red supergiant (RSG) progenitors for very nearby ($\lesssim$30 Mpc) Type IIP SNe \citep{2009ARA&A..47...63S}. 
These RSGs have masses ranging from 8 to 16 M$_\odot$. 
This mass range of Type IIP progenitors is further supported by comparing supernova observables to the predicted explosion outcome from realistic stellar models \citep[e.g.][]{2010MNRAS.408..827D,2012A&A...546A..28J,2013MNRAS.433.1745D,2015MNRAS.448.2482J}. 

The closest cousins to Type IIP SNe are the so-called Type II-Linear (IIL). 
Typical Type IIL SNe share similar spectroscopic properties as the Type IIP events; while their light curves decay linearly in magnitudes after peak. 
Despite early evidence of two distinct populations \citep[e.g.][]{2012ApJ...756L..30A}, recent studies of large samples have suggested that Type IIL and Type IIP SNe form a continuum in observed properties \citep[e.g.][]{2014ApJ...786...67A,2015ApJ...799..208S,2015A&A...582A...3G,2016arXiv160201446R}. 
We will therefore follow the recent literature and refer to these SNe collectively as Type II.
This category does not include the peculiar sub-classes of hydrogen-rich SNe and the Type IIb SNe, which have likely lost most of its hydrogen envelope and evolve to look like a Type I at late times.

The extended hydrogen-rich envelope dominates the ejecta mass of a Type II SN and plays an important role in shaping the early SN light curve.
However, the mass and structure of this envelope is strongly affected by poorly understood processes such as mass loss and binary evolution. 
In contrast, evolution of the core of the progenitor star is largely unaffected by these processes, as long as they occur post-main sequence.
The mass of the progenitor metal core is a sensitive probe of the main-sequence mass of the star \citep{2002RvMP...74.1015W,2007PhR...442..269W}.
Together with its pre-SN properties, the initial mass of a star constrains how it has evolved until the explosion. 

Nearby Type II SNe are valuable targets because they can be monitored for a long time after the explosion. 
When the outer ejecta have expanded and become optically thin, the core is revealed. 
During this nebular phase, emission lines form predominantly in the dense core region, with profiles shaped by the density and velocity distribution. 
Modeling of the nebular phase spectra has proved to be a powerful way to distinguish between progenitor models \citep{2011MNRAS.410.1739D,2012A&A...546A..28J}.

SN 2013ej exploded in the face-on spiral galaxy Messier 74 (M74, also known as NGC 628) in July 2013. 
It was spectroscopically classified as a young Type II SN shortly after the discovery \citep{2014MNRAS.438L.101V}.
Direct identification of a progenitor star has been suggested by \citet{2014MNRAS.439L..56F}. 
The mass of the progenitor was estimated to be between 8 and 15.5 {\msun}.
However, due to contamination from a nearby source, additional observations after the SN has faded are required to confirm the association and further constrain the progenitor properties.

Based on the temperature evolution during the first few weeks after shock breakout, \citet{2014MNRAS.438L.101V} derived a progenitor radius of 400 -- 600 {\rsun}.
This is consistent with the measured luminosity of the proposed RSG progenitor \citep{2014MNRAS.439L..56F} and a M-type supergiant stellar effective temperature.

Late time photometry has established that SN 2013ej ejected merely 0.02 {\msun} of radioactive {\ni} \citep{2015ApJ...806..160B,2015ApJ...807...59H,2016ApJ...822....6D}. 
\citet{2015ApJ...806..160B} used a semi-analytical model to derive an ejecta mass of 12 {\msun}, a progenitor radius of 450 {\rsun} and an explosion energy of $\sim2\times10^{51}$ erg.
A slightly smaller ejecta mass of around 10.6 {\msun} was estimated by hydrodynamical modeling in \citet{2015ApJ...807...59H}.
\citet{2016ApJ...817...22C} have used X-ray observations to infer the mass loss rate and derive a model-dependent zero-age main sequence (ZAMS) mass of around 14 {\msun} for the progenitor.
All these models are consistent with the current constraints from the direct detection of the progenitor star.

In this paper, we present the largest data set of multi-band optical and near-infrared (NIR) photometry and spectroscopy for SN 2013ej, acquired from 1 to 450 days after the explosion. 
Using the spectral synthesis models developed in \citet{2014MNRAS.439.3694J}, we take the first quantitative look at the nucleosynthesis of SN 2013ej.

Observations and data reduction procedures are presented in $\S$\ref{sec:obs}.
This is followed by a detailed look at the photometric evolution in $\S$\ref{sec:phot}.
We examine the early spectroscopic evolution of SN 2013ej in $\S$\ref{sec:spec_early};
then in $\S$\ref{sec:spec_late}, we inspect the late time spectra in light of synthesized spectral models from \citet{2014MNRAS.439.3694J}.
Implications of our results are discussed in $\S$\ref{sec:discussion} and we conclude in $\S$\ref{sec:conclusion}.

For analysis throughout the paper, we adopt a Galactic extinction of $E(B-V)$=0.06~mag ($R_V$=3.1) along the line of sight to the SN \citep{2011ApJ...737..103S} and zero reddening from the host galaxy. 
Negligible extinction is consistent with the non-detection of {\naid} absorption at the host redshift \citep{2014MNRAS.438L.101V}.
We use a host galaxy recession velocity of $657~\rm{km}~\rm{s}^{-1}$.
A range of distances (between 6.7 and 10.2 Mpc\footnote{\label{ned}\url{http://ned.ipac.caltech.edu/}}) have been estimated for the host galaxy M74.
Using our measured $I$-band magnitude ($12.50\pm0.05~\rm{mag}$ at day 50 after correction for extinction), photospheric velocity ($4450\pm50~\rm{km~s^{-1}}$ at day 50 measured from absorption minima of Fe II 5169\AA\ and 5018\AA)
and H$_{\rm 0}$ of $70~\rm{km}~\rm{s}^{-1}~\rm{Mpc}^{-1}$ \citep{2013ApJS..208...20B}, 
we estimate a distance of $10.2\pm0.9~\rm{Mpc}$ \citep[eq~(1) of ][]{2015A&A...580L..15P}.
Considering other distances derived from SN 2013ej \citep{2014AJ....148..107R,2016ApJ...822....6D},
we assume a mean distance of $9.7\pm0.5~\rm{Mpc}$, where the uncertainty is the standard deviation of the different measurements.

\section{Observations}\label{sec:obs}

\subsection{Optical Photometry}
The LCOGT network started an extensive monitoring campaign of SN 2013ej a few days after its discovery \citep{2014MNRAS.438L.101V}.
Multi-band ($UBVRIgriz$) photometric observations were carried out with nine 1-m telescopes. 
We measure aperture photometry with Source Extractor \citep{1996A&AS..117..393B}.
The $gri$-band observations are calibrated against the AAVSO Photometric All-Sky Survey\footnote{\url{http://www.aavso.org/apass}} DR7 catalog using field stars.
Calibration uncertainties are added to the statistical errors in quadrature.
For other optical bands, photometry of local reference stars are derived from observations of standard stars in photometric conditions. 

SN 2013ej was serendipitously detected by the 1-m telescope at the Lulin Observatory in Taiwan (LOT) when the field of M74 was observed as part of a training program at Taipei First Girls' High School \citep{2013ATel.5466....1L}. 
We measure aperture photometry with Source Extractor and apply the same calibration as for the LCOGT observations.
The discrepancy between magnitudes reported in \citet{2013ATel.5466....1L} and this paper is mainly due to differences in calibration catalogs being used.

SN 2013ej was also imaged at the two telescopes of the Mt. Ekar observing station of the INAF - Astronomical Observatory of Padova: 10 epochs were obtained using the Schmidt 67/92-cm Telescope equipped with an 
SBIG STL CCD with Kodak KAI-11000M (4049x2672 px) Dual Sensor; 1 epoch was obtained using the 1.82-m
Copernico Telescope equipped with AFOSC. Images were overscan, bias and flat-field corrected. Photometric measurements were obtained with the PSF-fitting technique, using the SNOoPY\footnote{SuperNOva PhotometrY, a package for SN photometry implemented in IRAF by E. Cappellaro; \\ \url{http://sngroup.oapd.inaf.it/snoopy.html}} package, and calibrated with reference to the magnitudes of the LCOGT reference stars.

On Aug 15, 2014, approximately one year after its discovery, SN 2013ej was imaged by ESO New Technology Telescope (NTT) with the ESO Faint Object Spectrograph and Camera (EFOSC2).
No APASS stars are covered by the EFOSC2 field-of-view. 
We perform aperture photometry with Source Extractor
and use identified point sources to tie the calibration to the LCOGT observations.

On the nights of Oct 19, 21 and 23 in 2014, SN 2013ej was imaged in the $V$-band by the X-shooter \citep{2011A&A...536A.105V} Acquisition and Guiding camera on the Very Large Telescope (VLT).
We calibrate the photometry relative to point sources identified in the NTT $V$-band image and take the weighted mean of the three observations.

All optical and NIR photometry data are listed in Table~\ref{tab:phot}.

\subsection{NUV Photometry}
SN 2013ej was observed by the Ultraviolet/Optical Telescope \citep[UVOT;][]{2005SSRv..120...95R} onboard the $SWIFT$ satellite from 2013 July 30.87 UT.
Near-UV photometry (in $uvw1$, $uvm2$ and $uvw2$-band) has been measured using an aperture of 3\arcsec\  following the approach of \citet{2009AJ....137.4517B}.

\subsection{Optical Spectroscopy}
Spectroscopic time series were obtained for SN 2013ej as part of the PESSTO\footnote{Public ESO Spectroscopic Survey of Transient Objects, \url{www.pessto.org/}} campaign. 
Spectra taken with NTT EFOSC2 were reduced with the PESSTO pipeline \citep{2015A&A...579A..40S}.

Early spectra of SN~2013ej were obtained with the robotic FLOYDS spectrograph mounted on the Faulkes Telescope South (FTS) at Siding Spring Observatory in Australia and the Faulkes Telescope North (FTN) at Haleakala Observatory in the U.S. state of Hawaii.
Most of these observations have been used in the analysis of \citet{2014MNRAS.438L.101V}. Observations that are not previously published are listed in Table~\ref{tab:spec_log}.

Some spectra of SN~2013ej were collected using the Wide Field Spectrograph \citep[\wifes;][]{dopita07, dopita10} on the Australian National University (ANU) 2.3-m telescope at Siding Spring Observatory in northern New South Wales, Australia.  WiFeS spectra were obtained using the B3000 and R3000 gratings, providing wavelength coverage from 3500 \AA\ to 9600 \AA\ with a resolution of 1.5 \AA\ and 2.5 \AA\ (all reported instrument resolutions are full width at half-maximum intensity, FWHM, of night sky lines) in the blue and red channels, respectively. Data cubes for WiFeS observations were produced using the PyWiFeS software \citep{pywifes}.

Spectra of SN 2013ej were obtained using the 1.82-m Copernico Telescope (+ AFOSC) and the 1.22-m Galileo Telescope of the Astrophysical Observatory of Asiago (Padua University) equipped with a B\&C spectrograph. AFOSC spectra were obtained using the grism\#4 (range 3500-8200 \AA; resolution $\sim$ 14 \AA) and the 
VPH6 (range 4700-10050 \AA; resolution 15 \AA). The B\&C spectra were obtained using the 300 lines/mm grating with a wavelength range 3400-7950 \AA\ 
and a typical resolution of 7 \AA. Spectroscopic data  reduction has been performed in a standard manner, using tasks available in the IRAF environment.

One spectrum was obtained on Aug 12, 2013 using Gemini Multi-Object Spectrograph (GMOS) on the Gemini South telescope, with the R400 grating and the 1.5 \arcsec\ longslit. Data were reduced using standard reduction tasks in IRAF.  

Additional nebular spectra were obtained with X-shooter \citep{2011A&A...536A.105V} on the VLT on Oct 19, 21 and 23, 2014. The data were reduced with the ESO X-shooter pipeline\footnote{http://www.eso.org/sci/software/pipelines/xshooter/} (version 2.5.2) and the Reflex interface. We use the median spectrum of the three epochs for our analysis. 

All spectra that are not yet public will be available on the Weizmann Interactive Supernova Data Repository\footnote{\url{http://wiserep.weizmann.ac.il}} \citep[WISeREP;][]{2012PASP..124..668Y} once the paper is published.

\subsection{NIR observations}
We secured near-IR ($JHK$) observations with the 60-cm Rapid-Eye-Mount (REM) telescope at the ESO La Silla Observatory (Chile) between 2013 August 3 and 2014 January 10. Each epoch consisted of a series of dithered images where each dither cycle comprised of five images. The integration time of a sub-image was 15 s in $J$ and $H$ band, and 10 s in $K$ band. 
The NIR images were reduced with standard routines in IRAF. For each dither cycle we constructed a sky image. After subtracting the sky from each image, these images were registered with \texttt{alipy} v2.0 and coadded.\footnote{\url{http://obswww.unige.ch/~tewes/alipy/}}. The world coordinate system was calibrated with the software package \texttt{astrometry.net} \citep{2010AJ....139.1782L}.
The photometry was done with Source Extractor. Once an instrumental magnitude was established, it was photometrically calibrated against the brightness of a number of field stars measured in a similar manner. 

NIR photometric and spectroscopic observations were obtained with the Son OF ISAAC (SOFI) at NTT, as part of the PESSTO monitoring.
SOFI data were reduced with the PESSTO pipeline \citep{2015A&A...579A..40S}.

All NIR photometry is calibrated to the 2MASS \citep{Skrutskie2006a} catalogue.

NIR spectra are also part of the X-shooter observation on Oct 21, 2014.

\begin{table}
\caption{Spectroscopic observations of SN 2013ej.}
\label{tab:spec_log}
\scriptsize
\begin{tabular*}{84.4mm}{@{\extracolsep{\fill}}cccc}
\hline
\multirow{2}{*}{UT Date} & \multirow{2}{*}{Phase\textsuperscript{a}} & Telescope & Wavelength\\
 & & +Instrument(+Setting) & Range (A) \\
\hline
13/07/31 &           +7 &            Asiago1.2m+BC+300tr &  3400 -- 7950  \\
13/07/31 &           +7 &                 ANU 2.3m+WiFeS &     3500 -- 9250  \\
13/08/01 &           +8 &            Asiago1.2m+BC+300tr &  3450 -- 7950 \\
13/08/02 &           +9 &            Asiago1.2m+BC+300tr &  3450 -- 7950  \\
13/08/03 &          +10 &            NTT3.6m+EFOSC2+gr11 & 3350 -- 7400  \\
13/08/03 &          +10 &            NTT3.6m+EFOSC2+gr16 & 6000 -- 10250 \\
13/08/03 &          +10 &       NTT3.6m+EFOSC2+gr11+gr16  & 3350 -- 10250  \\
13/08/03 &          +10 &            Asiago1.2m+BC+300tr & 3400 -- 7950  \\
13/08/04 &          +11 &       NTT3.6m+EFOSC2+gr11+gr16  & 3350 -- 10250 \\
13/08/07 &          +14 &            Asiago1.2m+BC+300tr & 3450 -- 7950  \\
13/08/08 &          +15 &            Asiago1.2m+BC+300tr  & 3400 -- 7950  \\
13/08/11 &          +18 &                 ANU 2.3m+WiFeS & 3500 -- 9250 \\
13/08/12 &          +19 &                 Gemini South+GMOS   & 4250 -- 8400   \\
13/08/13 &          +20 &                 Ekar1.8m+AFOSC & 3350 -- 8150 \\
13/08/14 &          +21 &       NTT3.6m+EFOSC2+gr11+gr16  & 3350 -- 9900 \\
13/08/19 &          +26 &                 ANU 2.3m+WiFeS  & 3500 -- 9250 \\
13/08/21 &          +28 &            Asiago1.2m+BC+300tr  & 3400 -- 7950  \\
13/08/22 &          +29 &            Asiago1.2m+BC+300tr  & 3450 -- 7950 \\
13/08/26 &          +33 &       NTT3.6m+EFOSC2+gr11+gr16  & 3350 -- 9900 \\
13/08/27 &          +34 &       FTS+FLOYDS  & 3200 -- 10000 \\
13/08/29 &          +36 &       NTT3.6m+EFOSC2+gr11+gr16  & 3350 -- 10250 \\
13/08/29 &          +36 &       FTS+FLOYDS  & 3200 -- 10000 \\
13/08/31 &          +38 &       FTS+FLOYDS  & 3200 -- 10000 \\
13/09/02 &          +40 &       FTS+FLOYDS  & 3200 -- 10000 \\
13/09/04 &          +42 &            Asiago1.2m+BC+300tr  & 3450 -- 7950 \\
13/09/04 &          +42 &       FTS+FLOYDS  & 3200 -- 10000 \\
13/09/08 &          +46 &       NTT3.6m+EFOSC2+gr11+gr16  & 3350 -- 9900  \\
13/09/12 &          +50 &       NTT3.6m+EFOSC2+gr11+gr16  & 3350 -- 10250 \\
13/09/19 &          +57 &                 ANU 2.3m+WiFeS  & 3500 -- 9250 \\
13/10/02 &          +70 &       NTT3.6m+EFOSC2+gr11+gr16  & 3350 -- 10250 \\
13/10/17 &          +85 &            Asiago1.2m+BC+300tr  & 3550 -- 7950  \\
13/10/24 &          +92 &                 ANU 2.3m+WiFeS & 3500 -- 9250 \\
13/10/26 &          +94 &       NTT3.6m+EFOSC2+gr11+gr16  & 3400 -- 10250 \\
13/11/23 &         +122 &       NTT3.6m+EFOSC2+gr11+gr16  & 3400 -- 10250  \\
13/12/10 &         +139 &                 Ekar1.8m+AFOSC  & 3500 -- 10050 \\
13/12/23 &         +152 &       NTT3.6m+EFOSC2+gr11+gr16  & 3400 -- 10250  \\
14/01/23 &         +183\textsuperscript{b} &            NTT3.6m+EFOSC2+gr11  & 3400 -- 7400 \\
14/01/30 &         +190\textsuperscript{b} &            NTT3.6m+EFOSC2+gr16  & 6000 -- 9950  \\
14/08/16 &         +388 &            NTT3.6m+EFOSC2+gr13  & 3650 -- 9250  \\
14/10/21\textsuperscript{c} &         +454 &                   VLT+X-shooter  & 3250 -- 24200  \\
\hline
13/08/16 &          +23 &                       NTT+SOFI+GB+GR  & 9350 -- 25250  \\
13/08/28 &          +35 &                       NTT+SOFI+GB+GR & 9400 -- 25250  \\
13/10/03 &          +71 &                       NTT+SOFI+GB+GR  & 9400 -- 25250  \\
13/10/25 &          +93 &                       NTT+SOFI +GB+GR & 9400 -- 25250  \\
13/12/09 &         +138 &                       NTT+SOFI+GB+GR & 9400 -- 25250  \\
\hline
\\
\multicolumn{4}{l}{\textsuperscript{a}\footnotesize{Phase relative to shock breakout date of Jul 24.0 (JD = 2456497.5).}}\\
\multicolumn{4}{l}{\textsuperscript{b}\footnotesize{Combined to construct a day 186 spectrum for analysis.}}\\
\multicolumn{4}{l}{\textsuperscript{c}\footnotesize{Median spectrum of data taken on Oct 19, 21 and 23.}}\\

\end{tabular*}
\end{table}

\section{Photometric Evolution}\label{sec:phot}
\begin{figure}
\includegraphics[width=84.4mm]{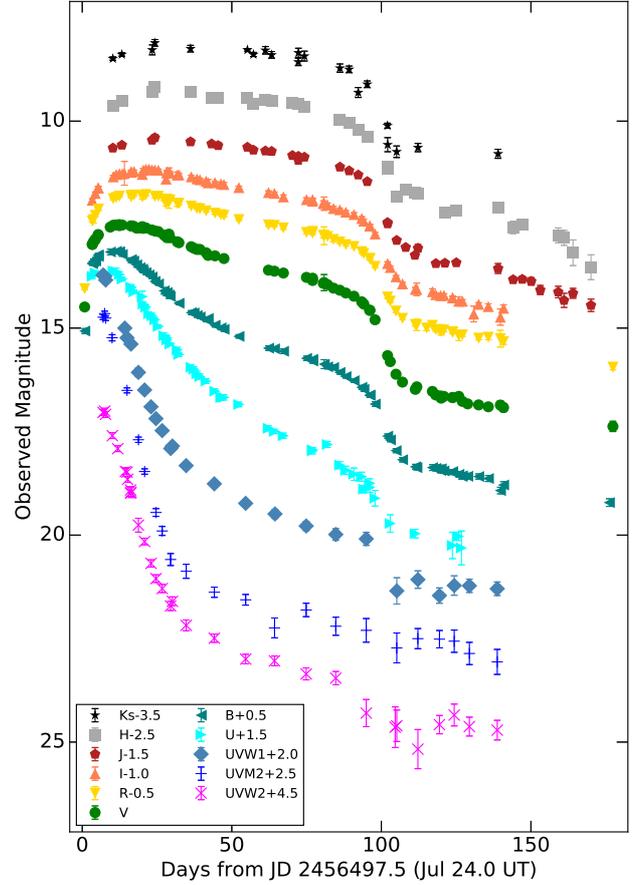}
\caption{Observed multi-band light curves of SN 2013ej in the first year (up to 177 days after explosion). Near-UV observations are from Swift UVOT. Optical observations are from LCOGT, INAF and LOT (first epoch). NIR observations are from REM and SOFI. Details of the optical and NIR measurements are listed in Table~\ref{tab:phot}.}
\label{fig:multi_lc}
\end{figure}

We first examine the photometric behavior of SN 2013ej. Figure \ref{fig:multi_lc} shows the observed multi-band light curves of the supernova in the first year.
Our study focuses on the optical and near-IR observations but we have included near-UV light curves to be complete.
SN 2013ej faded slowly between 50 and 100 days in all observed bands from near-UV to near-IR. 
Such an UV plateau is also identified by \citet{2015ApJ...806..160B}, and pointed out to be among the few well-observed cases \citep[see e.g. SN 2012aw][]{2013ApJ...764L..13B}.

\subsection{Time of Shock Breakout}\label{sec:early_lc}

\begin{figure}
\includegraphics[width=84mm]{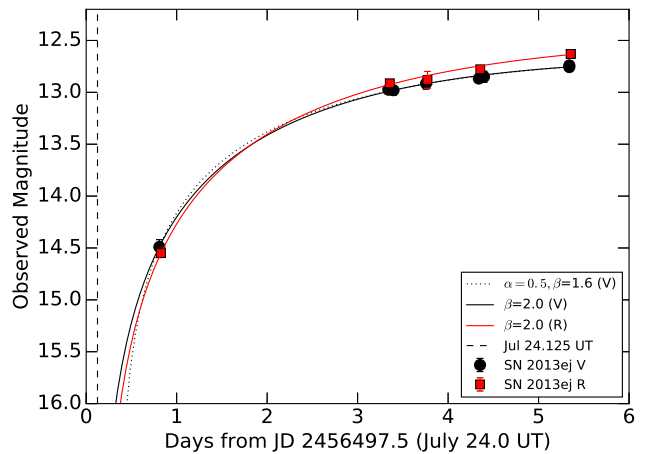}
\caption{Early light curves of SN 2013ej, compared to models described in $\S$~\ref{sec:early_lc}. Detection (without measured photometry) on Jul 24.125 UT is marked by a vertical dashed line.}
\label{fig:fit_lc}
\end{figure}

Shortly after the ``shock breakout'' of an SN, the emission is dominated by the expanding and cooling envelope.
Following \citet{2007ApJ...667..351W,2010APh....33...19C,2011ApJ...728...63R,2011ApJ...736...76R},
we model the early temporal evolution of the flux at a fixed wavelength as:

\begin{equation}\label{eq:t0}
f_{\lambda} = \frac{A}{e^{B (t-t_0)^{\alpha}}-1} (t-t_0)^{\beta}
\end{equation}

\noindent where $t_0$ is the time of explosion, $\alpha$ describes the temperature evolution and $\beta$ depends on the expansion of the photospheric radius.
\citet{2007ApJ...667..351W,2011ApJ...728...63R} estimate $\alpha$ to be around 0.5 and $\beta$ around 1.6,
for $t > t_{BO}$ when the photosphere lies outside the ``breakout shell''. 
For $t < t_{BO}$, \citet{2011ApJ...728...63R} suggest that the same temperature evolution should still be a valid approximation, while the photospheric radius may have a steeper dependence on $t$ (equivalent $\beta$ of up to 2).

For a large progenitor radius (few hundred {\rsun}, applicable for SN 2013ej), $t_{BO}$ is on the order of 1 day.
We choose a value of 2 for $\beta$ when fitting the $V$ and $R$-band light curves obtained in the first week and find the average best fit $t_0$ to be Jul 24.1 UT.
These models predict the SN to be fainter than 19th magnitude on Jul 24.125, therefore at odds with the reported detection (although without photometry) by C. Feliciano\footnote{\url{http://www.rochesterastronomy.org}}.
A smaller $\beta$ (e.g. 1.6) would yield an even later $t_0$ that is inconsistent with the Jul 24.125 detection.

It is likely that equation~\ref{eq:t0} does not adequately describe the early light curve.
\citet{2016ApJ...822....6D} estimated the date of shock breakout, using early unfiltered optical photometry, to be Jul $23.9\pm0.3$ UT.
For analysis in the rest of this paper, we adopt a shock-breakout date of Jul 24.0 UT (JD = 2456497.5), the mean of our estimate and that in \citet{2016ApJ...822....6D}.

\subsection{$V$-band Characteristics}
The $V$-band light curve of SN 2013ej exhibits all features identified in the $V$-band light curves of Type II SNe in \citet{2014ApJ...786...67A}. 
It first rapidly rose to a peak, then steadily declined. The decline rate slowed down about 50 days after explosion. 
After another 50 days or so, the light curve steepened and the brightness plunged by almost two magnitudes in 10 days. 
This leads to a linearly (in magnitude space) decaying tail until at least a year after explosion.

\begin{figure}
\includegraphics[width=84mm]{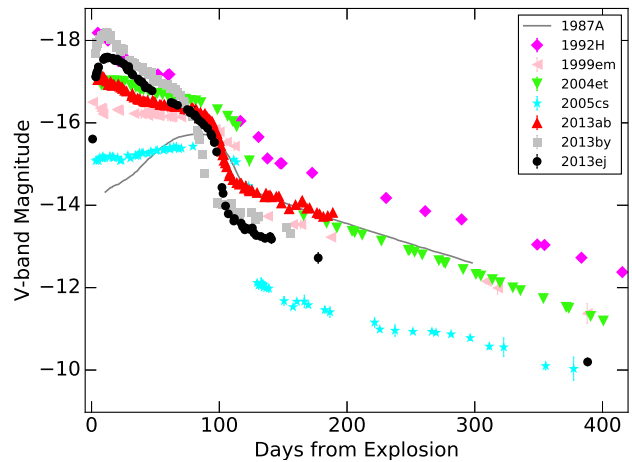}
\caption{$V$-band light curve of SN 2013ej, compared to selected Type II SNe.}
\label{fig:sne_v}
\end{figure}

SN 2013ej clearly declined faster than the prototypical Type IIP SNe (e.g. SN 1999em and SN 2004et) during the plateau phase (see Figure \ref{fig:sne_v}), which leads to its classification as a Type IIL SN \citep{2015ApJ...806..160B}. 
However, it shares an important characteristic with the IIPs, which is a rapid and significant brightness drop during transition from the plateau to the later tail phase. 
Such a dramatic transition suggests that the earlier light curve is powered by some energy source that ended abruptly. 
This is understood for Type IIP SNe as 
the initial release of thermal energy is regulated by hydrogen recombination
and the transition happens when the recombination wave reaches the bottom of the hydrogen-rich envelope. 
Given its spectroscopic similarity to other Type IIP SNe, the same explanation may apply to SN 2013ej.

The light curve shape of SN 2013ej is not unique. Several SNe in the literature have similar decline rates after the initial peak and some show evidence of a comparable transition to a tail. 
A recent well-observed example is SN 2013by \citep{2015MNRAS.448.2608V}. \citet{2015MNRAS.448.2608V} have also identified a few other fast declining SNe that exhibit the drop.
In fact, such transitions are observed in almost all Type IIL SNe that are monitored long enough but the transition times appear to be earlier than for the slow declining Type IIPs.

\begin{figure}
\includegraphics[width=84.4mm]{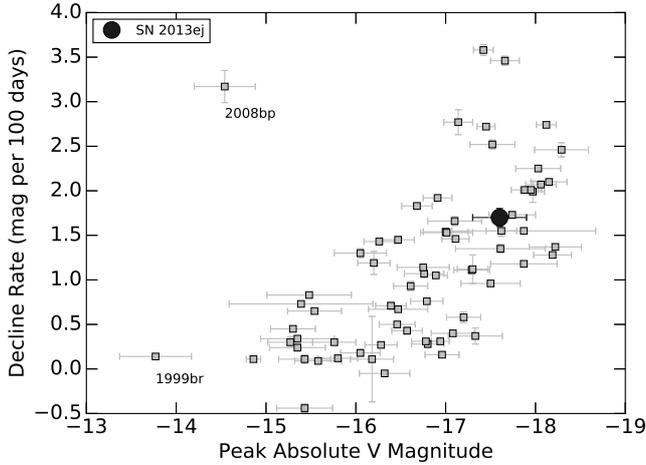}
\caption{$V$-band decay rate during the late plateau phase vs peak absolute $V$-band magnitude for SN 2013ej, compared to other Type II SNe from \citet{2014ApJ...786...67A}. 
The uncertainty in the peak absolute magnitude is dominated by the uncertainty in the distance of the SN.}
\label{fig:s2_max}
\end{figure}

We compare the light curve characteristics of SN 2013ej to the events presented in \citet{2014ApJ...786...67A}. 
SN 2013ej is relatively bright and has a relatively short plateau but all the measured parameters fall within the typical range. 
Figure \ref{fig:s2_max} shows that the decline rate during the plateau phase and the peak brightness of SN 2013ej follow the trend set by the majority of the Type II SNe.

\subsection{Color Evolution}

\begin{figure}
\includegraphics[width=84.4mm]{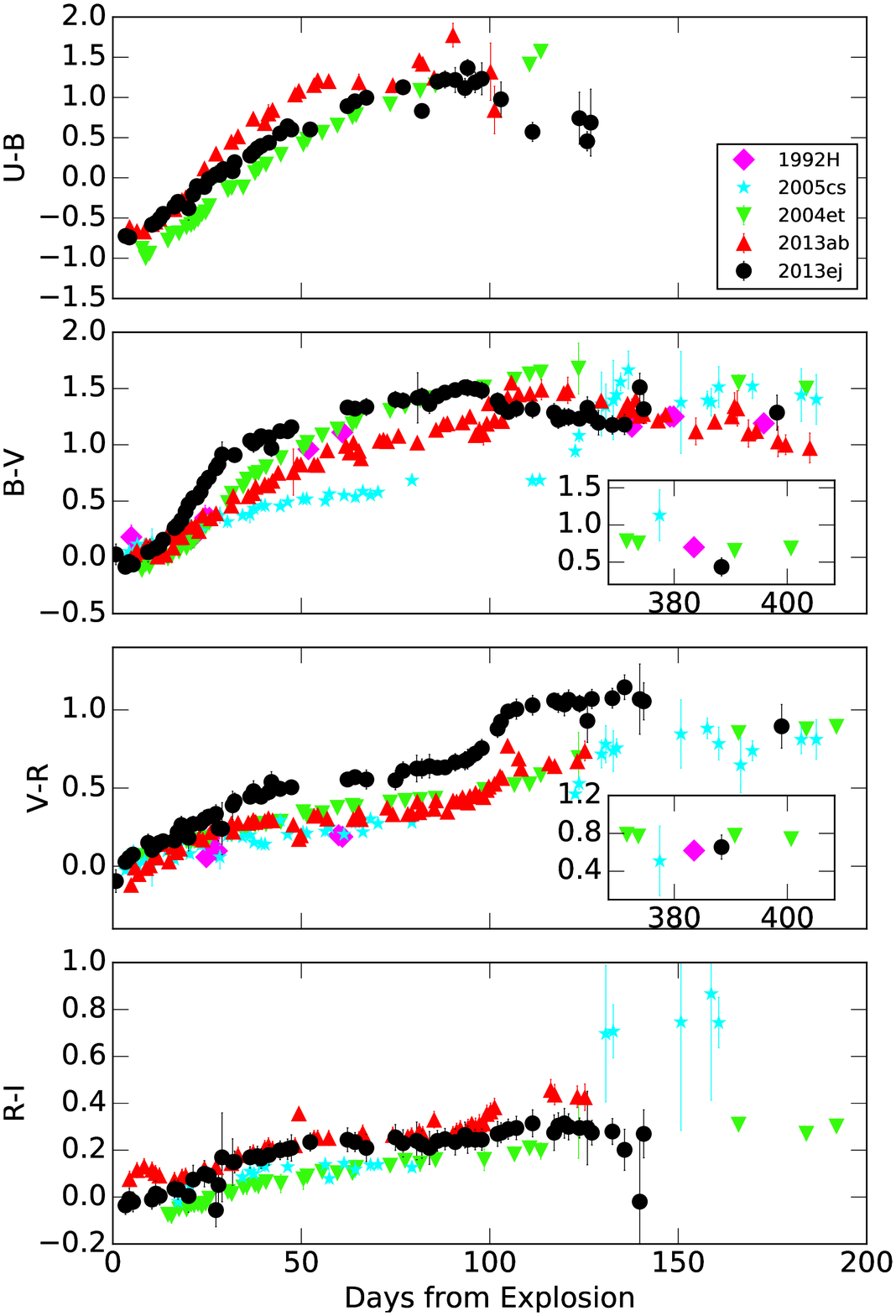}
\caption{Optical colour evolution of SN 2013ej, compared to selected Type II SNe. Data for other SNe are from \citet[][SN 1992H]{1996AJ....111.1286C}, \citet[][SN 2004et]{2006MNRAS.372.1315S}, \citet[][SN 2005cs]{2009MNRAS.394.2266P} and \citet[][SN 2013ab]{2015MNRAS.450.2373B}. Observed magnitudes are corrected for total extinctions estimated/used in the corresponding references.}
\label{fig:color_lc}
\end{figure}

\begin{figure}
\includegraphics[width=84.4mm]{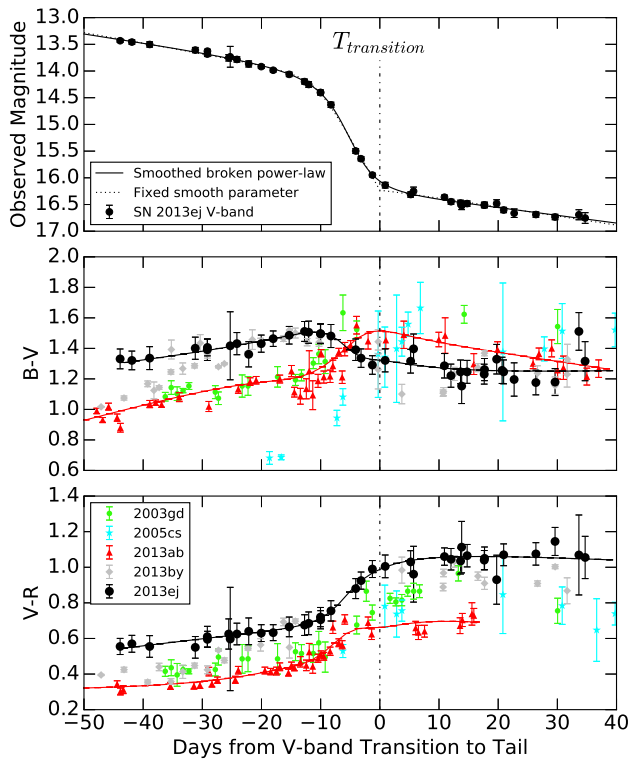}
\caption{Optical colour evolution of SN 2013ej and selected Type II SNe, around the transition to the tail phase. Top panel: examples of smoothed piecewise-linear fit (equation~\ref{eq:sbpl}, see text for details) for SN 2013ej. The transition time is defined as the second break time, when the smooth parameter $s_2$ is fixed to zero (no smoothing). Middle and bottom panels: colour evolutions relative to the transition times. For well-sampled light curves of SN 2013ab and SN 2013ej, we use the smoothed piecewise-linear model to interpolate the light curves and over-plot the continuous colour evolutions around the transition. Data for other SNe are from \citet[][SN 2003gd]{2005MNRAS.359..906H}, \citet[][SN 2005cs]{2009MNRAS.394.2266P}, \citet[][SN 2013ab]{2015MNRAS.450.2373B} and \citet[][SN 2013by]{2015MNRAS.448.2608V}. Observed magnitudes are corrected for total extinctions estimated/used in the corresponding references.}
\label{fig:color_transition}
\end{figure}

The optical colour curves of SN 2013ej are shown in Figure~\ref{fig:color_lc}, together with the colour curves of a few other Type II SNe from the literature. 
At early times, colours of all SNe grow redder as the ejecta envelopes cool, although the cooling rates vary from SN to SN. 
The $B-V$ colour of SN 2013ej changed quickly in the first month and appeared to be redder than the other SNe in the figure.
At late times, the colours of all SNe evolve slowly and are remarkably similar.
During the transition phase from plateau to tail, we observe different trends for different SNe, especially in the $B-V$ colour. 
A detailed comparison around this transition period is shown in Figure~\ref{fig:color_transition}, where we plot the colour evolutions relative to the end of the transition phase or the beginning of the decay tail. 

This ``transition time'' is chosen for two reasons. 
First, it is relatively easy to constrain for typical light curve shapes of Type II SNe (see below). 
Second, this transition time should roughly correspond to the time at which the photosphere reaches the helium core \citep{2011MNRAS.410.1739D}. 
The exact physical interpretation and the accuracy of the measurement is not important because the time is only used as a reference for qualitative comparison between SNe. 
Determination of this time does not depend on the explosion epoch, so an event such as SN 2003gd, which was only discovered near the end of the plateau phase, can be included in the comparison. 

To define this ``transition time'', we empirically model the $V$-band magnitude around the transition (between 50 and 200 days after the nominal explosion date) with a smoothed piecewise-linear function (or power-law in flux space):
\begin{equation}\label{eq:sbpl}
\begin{split}
m_v=a_0+a_1t+a_2\left(\frac{t-t_1}{2}+\sqrt{\frac{(t-t_1)^2}{4}+s_1}\right) \\
+a_3\left(\frac{t-t_2}{2}+\sqrt{\frac{(t-t_2)^2}{4}+s_2}\right)
\end{split}
\end{equation}
When $s_1=s_2=0$, the function reduces to piecewise-linear. The light curve shape around the first break time ($t_1$) varies from SN to SN, but is usually smooth (with large $s_1$). 
In contrast, the second break is often sharp. We therefore define the ``transition time'' as the second break time ($t_2$) when the smooth factor $s_2$ is fixed to zero.
Examples of the fit for SN 2013ej are shown in the top panel of Figure~\ref{fig:color_transition}. The transition time $T_{transition}$ is at day 106.1.
When this model is tightly constrained by data, we can also use it to interpolate the light curves across the transition, as in the case for SN 2013ej and SN 2013ab \citep{2015MNRAS.450.2373B}.

Just before the ``transition time'', the $B-V$ colour of SN 2013ej becomes bluer rapidly while SN 2013ab continues to become redder (but at a faster rate then the earlier plateau phase, see Figure~\ref{fig:color_transition}). 
SN 2003gd behaves similarly to SN 2013ab, as does the sub-luminous SN 2005cs, albeit with a much more dramatic change in its absolute $B-V$ colour. 
SN 2013by, another fast decaying Type II SN, appears to follow the same trend as SN 2013ej. 

The few $U$-band detections during and after the transition suggest the $U-B$ colour of SN 2013ej also becomes bluer.
Unfortunately, no spectrum is available during this time to see if the evolution is due to change in continuum or spectral features.
Even well-sampled multi-band light curves are rare in this small time window ($\sim$10 days) around the transition. 
Extensively-studied Type II SNe,  such as SN 2004et and SN 1999em, have poor photometric coverage during the transition phase. 
It remains to be seen whether the trend of colour evolution correlates with other light curve characteristics.

\subsection{Bolometric Light Curve}\label{sec:bolo}

We construct a pseudo-bolometric light curve for SN 2013ej using our optical and NIR observations. 
This is done in two ways. 
First, we estimate the spectral energy distributions (SED) with $UBgVrRiIJHK$ photometry from 10 to 170 days after explosion (range covered by $J$-band). 
Well-sampled light curves are interpolated daily with smooth cubic splines before the end of the plateau. 
After the plateau, curvature in the light curve is reduced and larger gaps exist, and we use smooth linear spline fits in magnitude space.
The SEDs constructed from the interpolated photometry are then integrated from 330 to 2400 nm, assuming flux densities go to zero linearly towards the boundaries. 
We have not included flux at shorter wavelengths in our calculation. 
Contribution from near-UV is significant in the first month \citep[almost 50\% at day 12, see Figure~7 in][]{2015ApJ...806..160B} but declines rapidly over time and becomes negligible (less than 4\% at day 40).

Alternatively, we measure fluxes in the same wavelength range (330 to 2400 nm) by integrating the photometry-calibrated spectra. 
For each NIR spectrum, we have at least one optical spectrum taken within 2 days. 
Each pair of NIR and optical spectra is calibrated by photometry to a common epoch assuming negligible spectral evolution. 

\begin{figure}
\includegraphics[width=84.4mm]{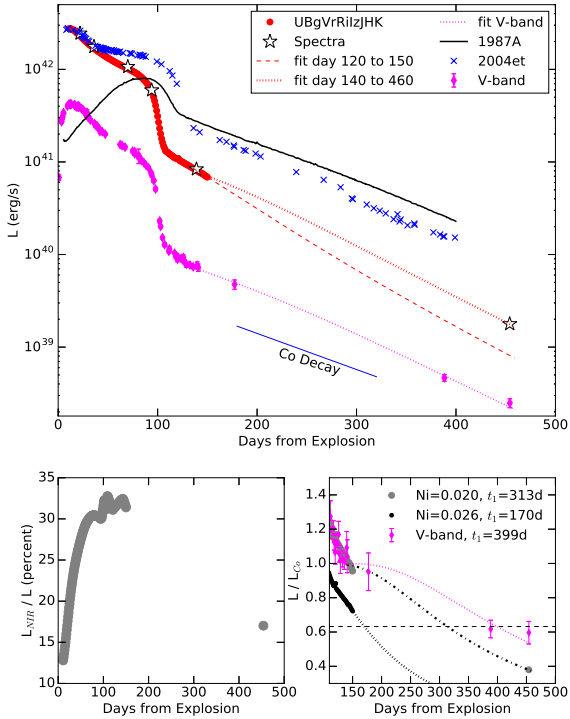}
\caption{Bolometric light curve of SN 2013ej. 
Upper panel: optical-NIR pseudo-bolometric light curves constructed from photometry or spectra, compared to $V$-band flux and bolometric light curves of SN 1987A and SN 2004et \citep{2010MNRAS.404..981M}.
The $V$-band and bolometric light curves are fitted by a {\co} decay model considering $\gamma$-ray leakage.
Bottom left: NIR contributions to the pseudo-bolometric light curve.
Bottom right: ratio between measured bolometric flux and {\co} decay model with full $\gamma$-ray trapping.
Best fit light curve models are plotted as thick dotted line (for early bolometric tail), dash-dotted line (for late bolometric tail) and thin dotted line for V-band. 
The horizontal dashed line corresponds to the fraction of trapped $\gamma$-ray energy when the optical depth is one.
}
\label{fig:bolo_lc}
\end{figure}

Pesudo-bolometric fluxes obtained with the two methods agree well (see Figure~\ref{fig:bolo_lc}). 
After day 200, photometry coverage becomes sparse and does not constrain the SED in NIR.
We integrate the X-shooter spectrum at day 454 to obtain the late bolometric flux.
Only $V$-band photometry is available at this epoch, so we do not account for uncertainty due to the relative calibration between optical and NIR.
We plot the $V$-band flux in Figure~\ref{fig:bolo_lc} for comparison.

We also estimate the fraction of flux above 1000 nm by integrating the SEDs from 1000 to 2400 nm (lower left panel of Figure~\ref{fig:bolo_lc}).
This fraction increases steadily as the ejecta cool and peaks at around 33 percent at the end of the plateau.
It is not clear whether the flat NIR light curves between 120 and 140 days (see also Figure~\ref{fig:multi_lc}) are physically real.
Overall, contribution from NIR decreases slowly during the decay tail, as for other Type II SNe \citep[e.g.][however the values are not directly comparable to this work because a different method is used]{2010MNRAS.404..981M}. 

After the plateau, the dominating power source is believed to be radioactive {\co}, a decay product of {\ni}.
Bolometric flux can thus be used to estimate the {\ni} production in the SN explosion.
If the $\gamma$-rays are fully trapped, the bolometric evolution should follow the decay of {\co}. 
However, the light curves of SN 2013ej decayed faster than expected from {\co} decay. 
This is also noted by \citet{2015ApJ...806..160B,2015ApJ...807...59H,2016ApJ...822....6D}.

To account for $\gamma$-ray leakage, we use the following model to estimate the {\ni} mass:

\begin{equation}\label{eq:co_leak}
L = 1.41\times 10^{43}~m_\mathrm{Ni}~(e^{-t/t_\mathrm{Co}}-e^{-t/t_\mathrm{Ni}})~(1-e^{-t_1^2/t^2})~\mathrm{erg}~\rm{s}^{-1}
\end{equation}

\noindent where $L$ is the total luminosity; $m_\mathrm{Ni}$ is the {\ni} mass in units of {\msun} ; $t$ is the time since explosion; 
$t_\mathrm{Co}$ and $t_\mathrm{Ni}$ are the e-folding times of {\co} and {\ni}, taken to be 111.4 and 8.8 days respectively; and $t_1$ is the characteristic time when the optical depth for $\gamma$-rays reaches one.
The $t^{-2}$ dependence for optical depth is due to homologous expansion \citep{1997ApJ...491..375C}.

For a spherical envelope with uniform density, $t_1$ depend on the mass $M_{env}$ and kinetic energy $E_k$ of the envelope:

\begin{equation}\label{eq:t1}
t_1 = \sqrt{\frac{9}{40\pi}~\kappa_{\gamma}~\frac{M_{env}^2}{E_k}}
\end{equation}

\noindent where $\kappa_{\gamma}$ is the $\gamma$-ray opacity \citep[assumed to be 0.033 $\mathrm{cm}^2\mathrm{g}^{-1}$;][]{1989ApJ...346..395W,2014ApJ...792...10S},
$E_k = (3/10)M_{env}v^2$ and $v$ is the characteristic ejecta velocity.
For SN 1987A, $t_1$ is estimated to be 530 days and the corresponding ejecta mass is 13 {\msun} \citep[see Fig 4.2 in][]{2011PhDT........90J}.

When {\ni} is mixed into the ejecta, $\gamma$-ray trapping is less efficient compared to a centralized {\ni} distribution. 
The effective $t_1$ can be evaluated using this equation as the time when the trapped fraction reaches 63.2\%.
The ejecta mass evaluated using equation~\ref{eq:t1} will be a lower limit.

Using the pseudo-bolometric light curve between 120 and 150 days, we estimate the {\ni} mass to be 0.026 {\msun} and $t_1$ to be 170 days.
\citet{2015ApJ...806..160B} use a similar formulation for the {\ni} decay component in their model and estimate an equivalent $t_1$ of 173 days.
However, this model underestimated the flux at day 454 by a factor of 3.
Yet, we do not see any abnormal features in the day 454 spectrum that can account for this additional flux.

Using equation~\ref{eq:co_leak} to fit the bolometric flux between 140 and 454 days, we obtain a {\ni} mass of 0.020 {\msun} and a $t_1$ of 313 days (see also lower right panel of Figure~\ref{fig:bolo_lc}).
This model is justified if residual radiation from the envelope contributes to the luminosity shortly after the plateau phase.
We also measure a $t_1$ of 399 days from fitting the $V$-band light curve. As shown in Figure~4 of \citet{2012A&A...546A..28J}, 
the fraction of light emerging in $V$-band stays constant up till 400 days in both models and in the well-studied SN 2004et.
The larger $t_1$ (> 300 days) is therefore preferred.

Considering both possibilities, we estimate the {\ni} mass produced in SN 2013ej to be $0.023\pm0.003$ {\msun}.
For a typical kinetic energy of $10^{51}$ erg, the ejecta mass is 5--9 {\msun}  (lower value corresponds to smaller $t_1$).
The mass estimate scales with $\sqrt{E_k}$.
In other words, small $t_1$ (significant $\gamma$-ray leakage) suggests small envelope mass or large kinetic energy.

\section{Early Spectral Evolution}\label{sec:spec_early}

\subsection{Optical spectra}

\begin{figure}
\includegraphics[width=84.4mm]{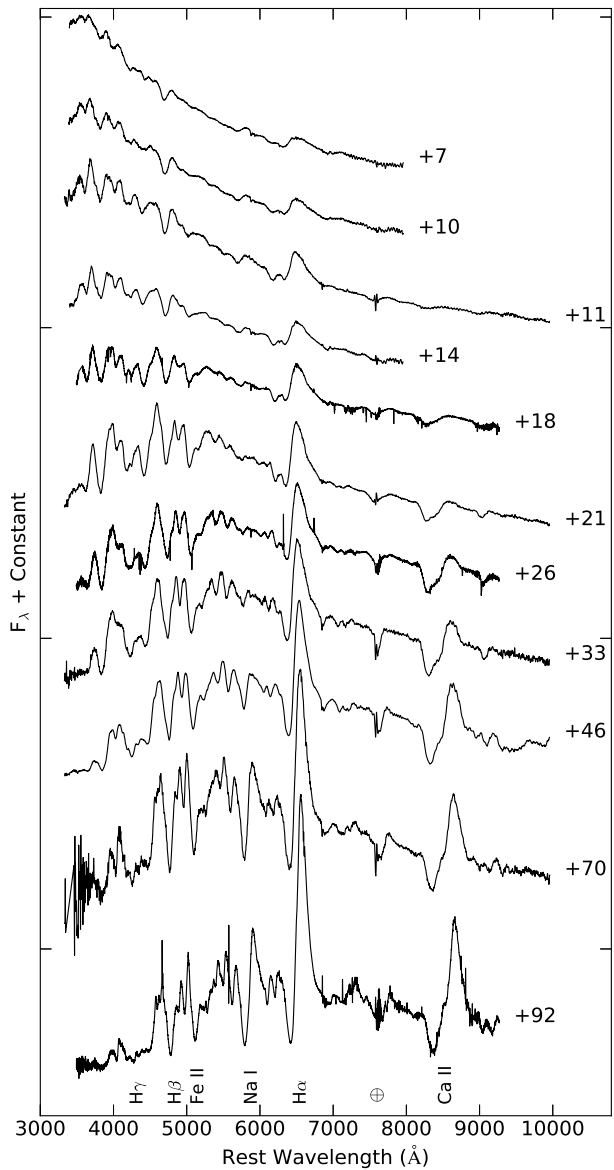}
\caption{Selected spectra of SN 2013ej during photospheric phase.}
\label{fig:spec_series1}
\end{figure}

Figure \ref{fig:spec_series1} shows how the optical spectra of SN 2013ej evolved during the ``plateau'' phase. The continuum starts hot and only hydrogen and perhaps also helium P-Cygni features are visible early on. As the photosphere cools down more lines appear and grow stronger. The early evolution was discussed in \citet{2014MNRAS.438L.101V}  and modeled to constrain the size of the progenitor. At 20 days after shock break out, {\naid} starts to appear and grows progressively stronger. Before the transition phase at approximately 100 days, \hi, \feii, {\nai} and {\caii} dominate the spectra.

\begin{figure}
\includegraphics[width=84.4mm]{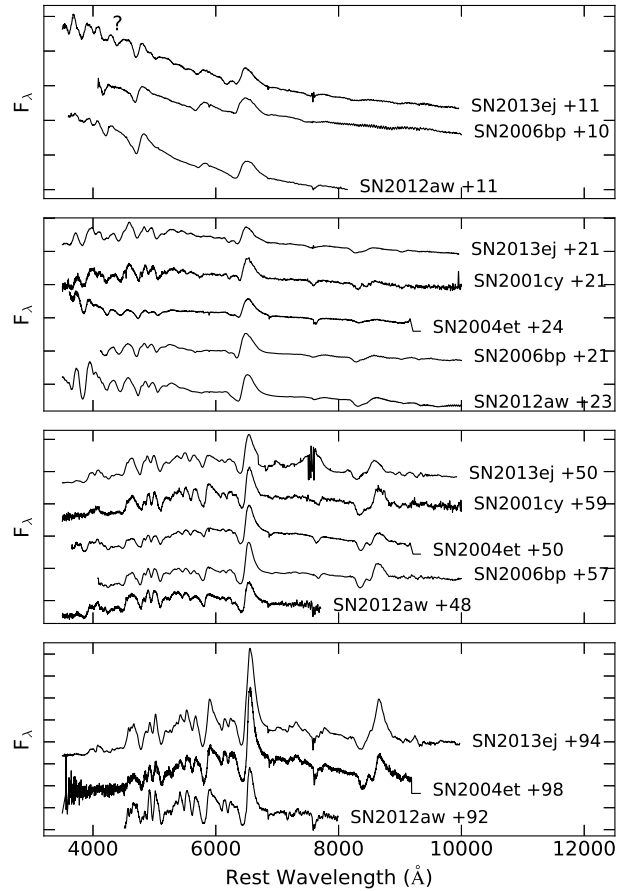}
\caption{Spectra of SN 2013ej compared to selected Type II SNe during the photospheric phase. Comparison spectra are from \citet{2006MNRAS.372.1315S, 2007ApJ...666.1093Q,2014ApJ...787..139D}.}
\label{fig:spec_comp}
\end{figure}

We compare SN 2013ej to a few other Type II SNe during the photospheric phase in Figure~\ref{fig:spec_comp}.  Qualitatively, the spectra of Type II SNe become indistinguishable after about 50 days from explosion. At early times, several differences are noted among the group. Around day 10, a clear absorption feature around 4400 \AA\ is only visible in the spectrum of SN 2013ej. 
The absorption on the blue side of {\ha} is identified as {\siii}, but not high velocity hydrogen, in both \citet{2014MNRAS.438L.101V} and \citet{2015ApJ...806..160B}.

Shallow absorption in the $H_{\alpha}$ profile or existence of an additional component (that fills in the absorption) has been noted to occur in fast decaying Type II SNe (e.g. Type IIL SN 2001cy). \citet{2014ApJ...786L..15G} have found that the ratio of absorption to emission of $H_{\alpha}$ correlates with a number of photometric properties. Fast decliners tend to have smaller absorption to emission ratios.

Photospheric velocities of SN 2013ej are measured from the Fe~II $\lambda$ 5018 and $\lambda$ 5159 absorption minima. The evolution (from $\sim$10000 to 3000 $\rm{km}~\rm{s}^{-1}$ just before the end of the plateau) is marginally higher but similar to other Type II SNe such as SN 2004et and SN 2012aw \citep[see e.g. Figure~17 in][]{2015ApJ...806..160B}.

\subsection{NIR spectra}

\begin{figure}
\includegraphics[width=84.4mm]{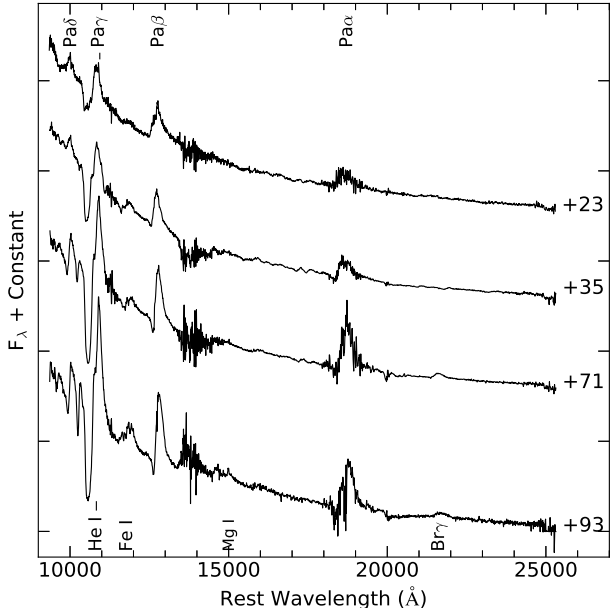}
\caption{NIR spectral series of SN 2013ej during photospheric phase.}
\label{fig:spec_series_nir}
\end{figure}

NIR spectra of SN 2013ej are dominated by the Paschen series and {\hei} 1.083 $\rm \mu$m (Figure~\ref{fig:spec_series_nir}).
In addition, Brackett $\gamma$ is clearly visible from day 70 onwards;
The emission complex between 1.15 and 1.2 $\rm \mu$m, likely due to Fe I, grows progressively stronger overtime;
\mgi\ 1.504 $\rm \mu$m is probably detected at day 93.

\section{Late-time Spectral Evolution}\label{sec:spec_late}

\subsection{Optical spectra}

\begin{figure}
\includegraphics[width=84.4mm]{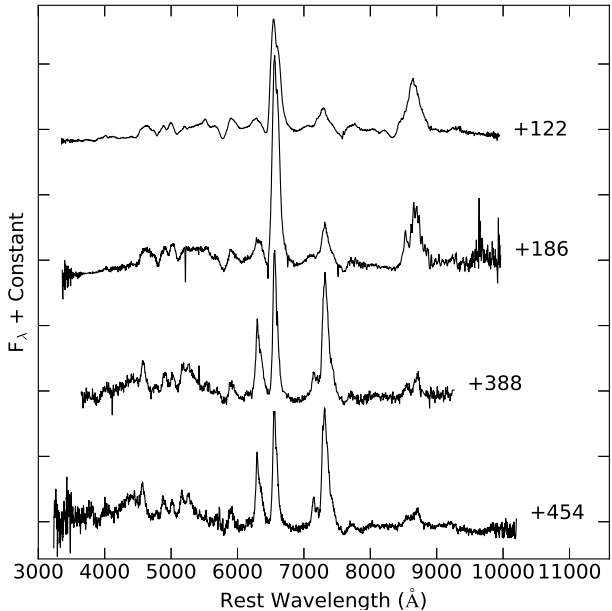}
\caption{Selected spectra of SN 2013ej during its tail decay phase. The spectrum at day 454 has been smoothed.}
\label{fig:spec_series2}
\end{figure}

\begin{figure*}
\begin{minipage}{\textwidth}
\includegraphics[width=180mm]{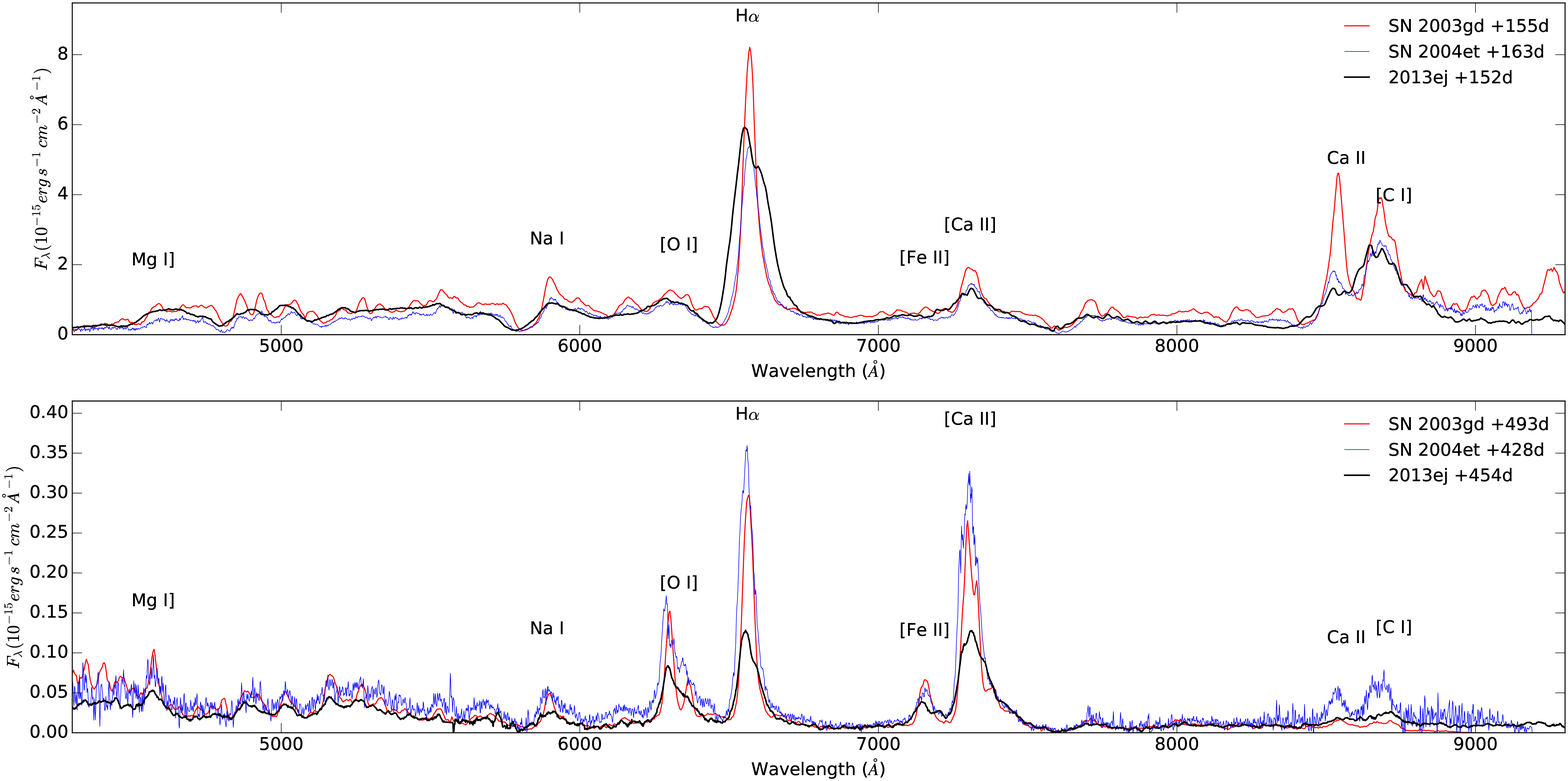}
\caption{Optical spectra of SN 2013ej compared to other SNe at day $\sim$150 and $\sim$450. Spectra of SN 2013ej are de-reddened and flux calibrated using multi-band photometry interpolated or extrapolated to those epochs. Optical spectra of SN 2003gd \citep{2005MNRAS.359..906H} and SN 2004et \citep{2006MNRAS.372.1315S} at similar phases are rescaled by their ratio of {\ni} mass and plotted for comparison. The flux of SN 2004et is rescaled to the distance of M74.}
\label{fig:neb_other}
\end{minipage}
\end{figure*}

After the ``plateau'' phase, the continuum luminosity plunged and the spectra became dominated by emission features. 
Over time, the ejecta continue to expand and expose deeper and denser layers of the inner core.
As seen in Figure~\ref{fig:spec_series2}, {\oi} $\lambda\lambda$6300, 6364 and [\caii] $\lambda\lambda$7291, 7323 grow progressively stronger;
while {\ha} becomes weaker.

The {\ha} emission profiles of SN 2013ej appear notably broad and asymmetric.
This is illustrated in Figure~\ref{fig:neb_other} where SN 2013ej is compared to two nearby Type II SNe.
SN 2003gd exploded in the same host galaxy as SN 2013ej and has a slightly smaller {\ni} mass of 0.015 {\msun} \citep{2005MNRAS.359..906H}.
SN 2004et is closer at 5.5 Mpc and produced 0.062 {\msun} of {\ni} \citep{2012A&A...546A..28J}.
We rescale the spectra of SN 2003gd and SN 2004et by the ratios of {\ni} mass and to the distance of M74.
Scaling with {\ni} mass is justified as the spectral shapes of SN 2003gd and SN 2004et are remarkably similar despite of the difference in {\ni} mass.

At around 150 days after explosion, emission lines of SN 2013ej are considerably broader and more blended. 
The difference diminishes over time. 
At around 450 days after explosion, the optical spectrum of SN 2013ej becomes similar but fainter at most wavelengths compared to the scaled spectrum of SN 2004et.
The difference may be explained by a smaller fraction of $\gamma$-rays trapped in SN 2013ej.

Among the three events, SN 2003gd has the smallest amount of {\ni} and the narrowest emission features.
Besides the difference in the line widths, noticeable differences exist around {\ha}, [\caii] $\lambda\lambda$7291, 7323 and {\caii} NIR triplet emissions.
Both SN 2003gd and SN 2004et show stronger {\ha} and [\caii] $\lambda\lambda$7291, 7323 than SN 2013ej.
SN 2003gd displays a particularly strong {\caii} $\lambda$8542 at day $\sim$150, but the triplet feature fades and becomes similarly weak as for SN 2013ej by day $\sim$450.
All three features, {\ha}, [\caii] $\lambda\lambda$7291, 7323 and {\caii} NIR, are formed mainly in the hydrogen-rich zone \citep{1993ApJ...405..730L,1998ApJ...497..431K,2012A&A...546A..28J}.
The diversity therefore suggests variation of the envelope properties.

\subsection{Emission line profiles}\label{sec:neb_profiles}

\begin{figure}
\includegraphics[width=84.mm]{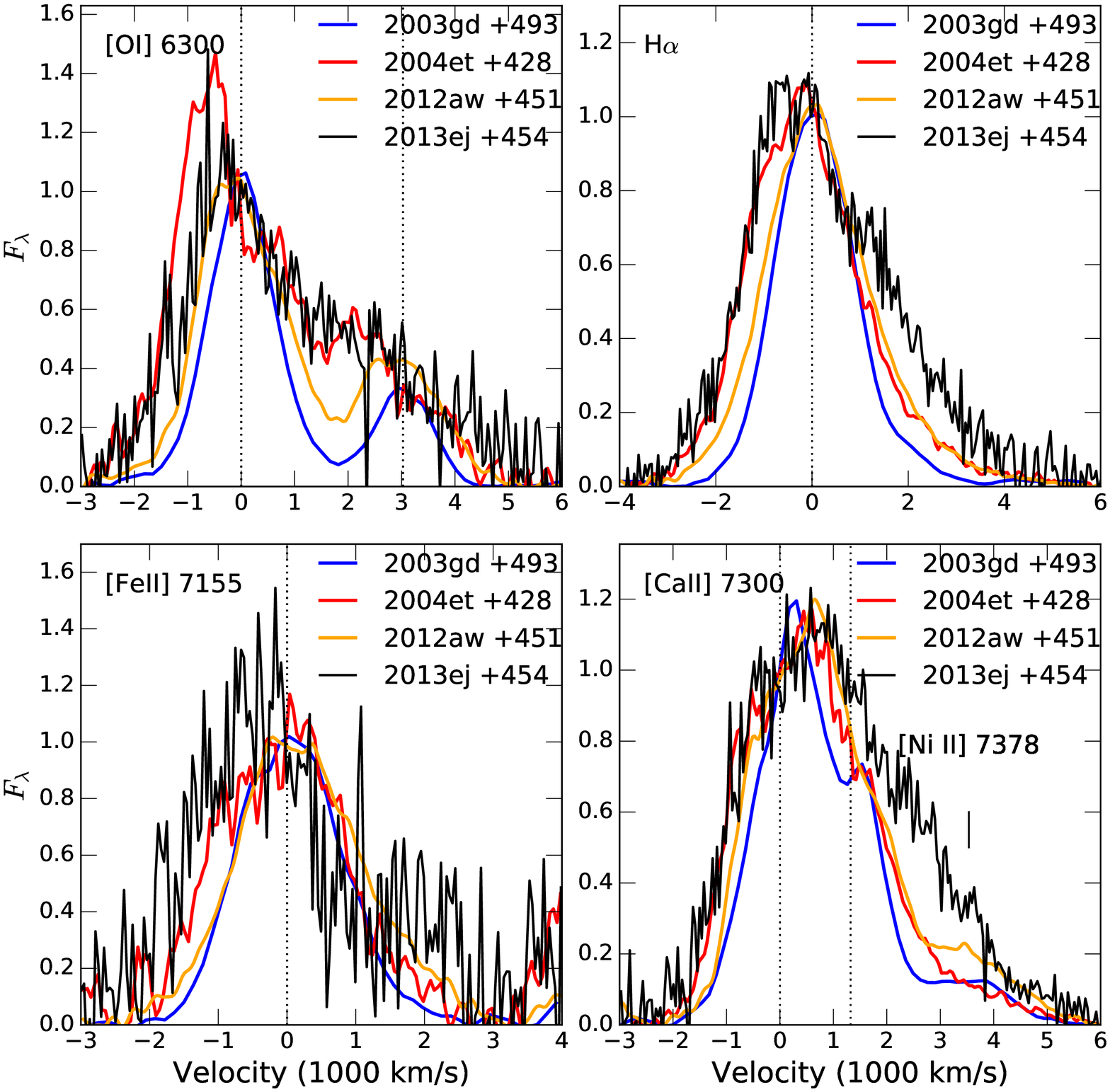}
\caption{Observed profiles of selected emission lines for SN 2013ej at day 454, compared to lines from other Type II SNe at similar epochs \citep{2005MNRAS.359..906H,2006MNRAS.372.1315S,2014MNRAS.439.3694J}. The lines are plotted in velocity space, relative to the central wavelength or the bluer component in a doublet. Vertical dotted lines mark the wavelengths of the single or doublet feature. The profiles are continuum subtracted and normalized. The continuum is estimated for each feature by fitting a straight line through the first minima on either side. 
In the lower right panel, the location of the [Ni II] $\lambda$7378 line is marked.}
\label{fig:lines_400}
\end{figure}

We further compare the profiles of the main emission features in Figure~\ref{fig:lines_400}. 
Using a sample of Type IIP SNe including SN 2003gd and SN 2004et, \citet{2012MNRAS.420.3451M} suggested a correlation between the {\ni} mass and the FWHM of the {\ha} emission lines. 
SN 2012aw, with a similar {\ni} mass estimate as SN 2004et \citep[0.06{\msun},][]{2014MNRAS.439.3694J}, appears to have emission lines that are wider than SN 2003gd and somewhat narrower than SN 2004et.
SN 2013ej, however, has lines at least as broad as those of SN 2004et (which produces 2 times more {\ni}).
The {\oi} $\lambda\lambda$6300, 6364 lines are similarly blended as SN 2004et.
Red wings of {\ha} and [\caii] $\lambda\lambda$7291, 7323 appear even more extended for SN 2013ej than for SN 2004et.

Resemblance between the {\ha} and [\caii] $\lambda\lambda$7291, 7323 profiles suggests that the broadness is intrinsic, 
not due to contamination from nearby features.
As seen in the lower right panel of Figure~\ref{fig:lines_400}, emission from [\niii] $\lambda$7378 is discernible for other objects,
but not for SN 2013ej.

\begin{figure}
\includegraphics[width=84.mm]{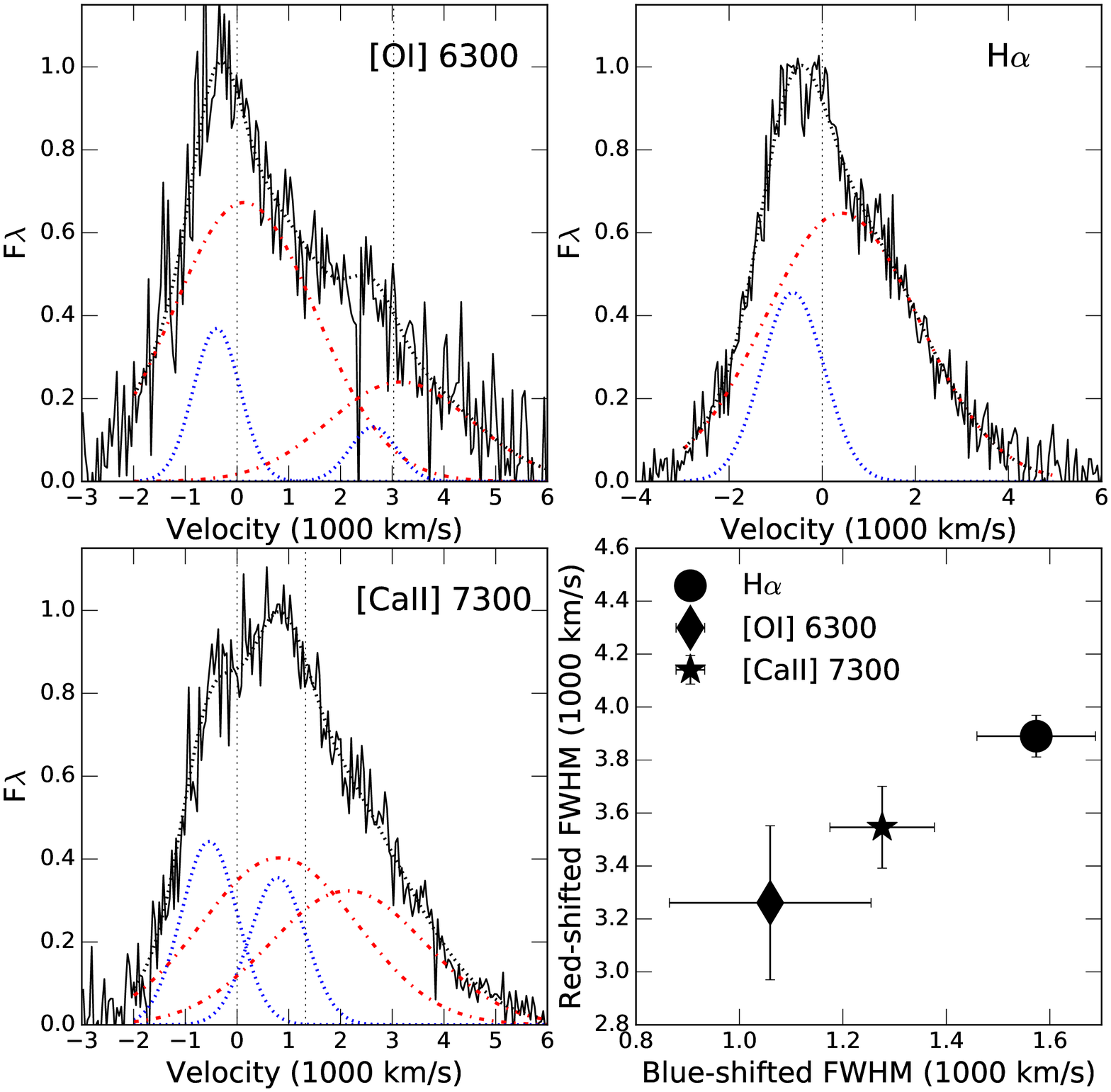}
\caption{Profiles of the [O I],  {\ha} and [Ca II] lines at day 454 are well fit by combinations of red-shifted and blue-shifted gaussian components.
Doublet features are modeled as two gaussians with the same width and fixed line ratios. 
FWHMs of the two components are plotted in the lower right panel, with 1$\sigma$ statistical errors.}
\label{fig:lines_multi}
\end{figure}

In Figure~\ref{fig:lines_multi},
we show that the broad and blended {\oi} $\lambda\lambda$6300, 6364, {\ha} and [\caii] $\lambda\lambda$7291, 7323 features
can be explained by two emission components, one blue-shifted and one red-shifted.
The profile parameters are estimated by least-squares minimization of the residuals.
When fitting the doublet features, we allow the intensity ratio of the two lines to vary but require this ratio to be the same for both components.
We also require that the doublet lines have the same width in the same component.
Both components have smallest width in {\oi} and largest width in {\ha}, consistent with where the lines are formed.
The {\oi} doublet has as an intensity ratio $I(6300)/I(6364)$ of $2.8\pm0.4$, indicating the line is optically thin at this epoch (the ratio is 1 in the optically thick limit and 3 in the optically thin limit).
No contribution from [\niii] $\lambda$7378 is required for the red wing of the [\caii] profile.

\subsection{Evolution of {\ha} emission}\label{sec:ha}

\begin{figure}
\includegraphics[width=84.4mm]{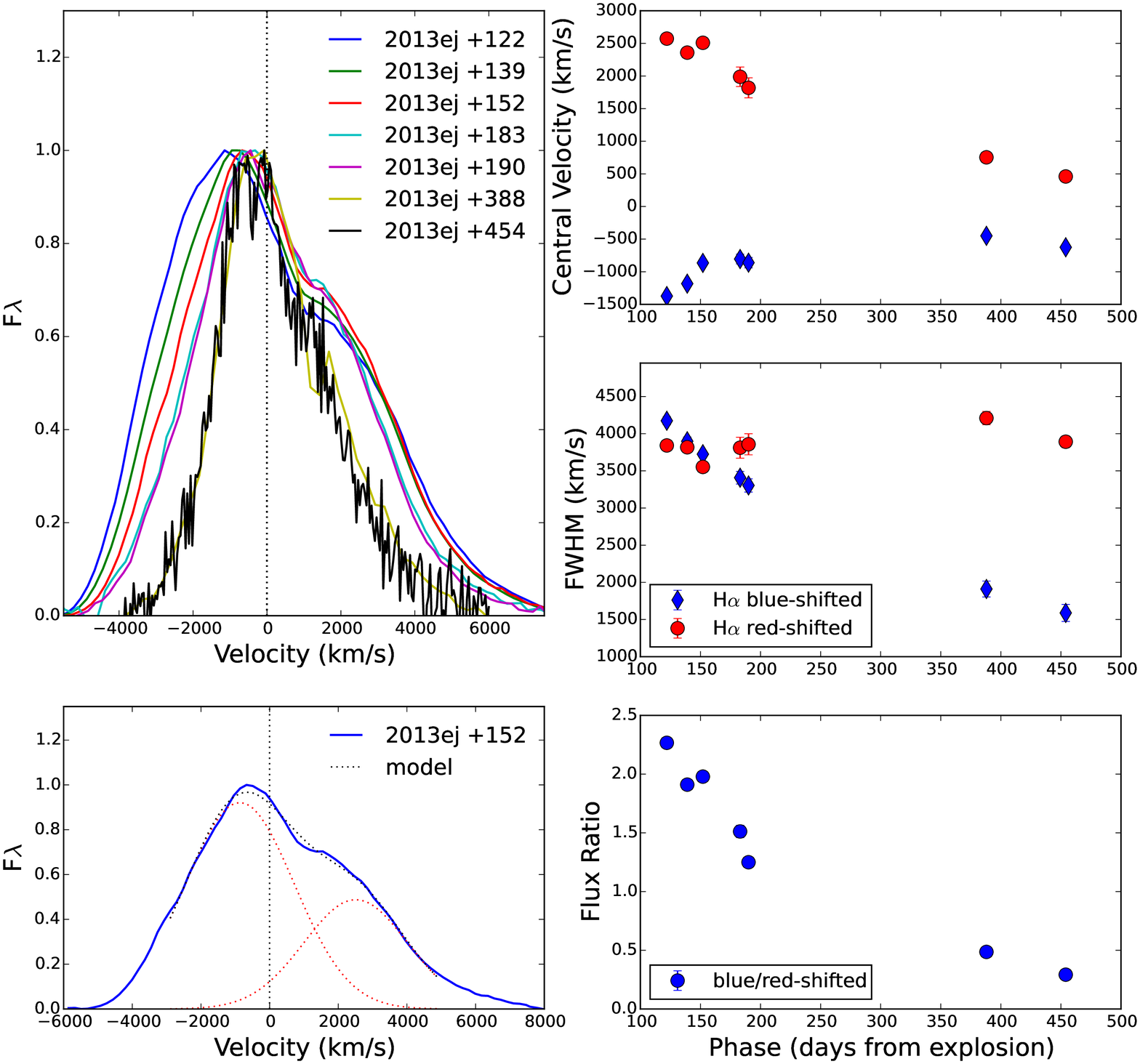}
\caption{Evolution of {\ha} profile from 122 to 454 days. Upper left: normalized {\ha} profiles. Lower left: example fit with two gaussian components.
Upper right: evolution of the central velocities of the two gaussian components. 
Middle right: evolution of the FWHM of the two gaussian components.
Lower right: evolution of the ratio of the blue-shifted and the red-shifted components.}
\label{fig:halpha}
\end{figure}

Asymmetric {\ha} profiles have been observed for a few Type II SNe and suggested to result from CSM interaction, asymmetry in the line-emitting region \citep{2002PASP..114...35L} 
or bipolar {\ni} distribution enclosed in a spherical envelope \citep{2006AstL...32..739C}.
Similar to Figure~\ref{fig:lines_multi}, we model the {\ha} profile with two components and inspect the temporal evolution of the two components in Figure~\ref{fig:halpha}.

The blue-shifted component has a steadily declining FWHM but a slow shift in central velocity.
The red-shifted component has a roughly constant (and possibly increasing) FWHM but faster evolution in central velocity.
Relative strength of the red-shifted component increases overtime,
which is opposite to the expected trend caused by dust formation.
Qualitatively, the {\oi} $\lambda\lambda$6300, 6364 feature evolves similarly, although it is weaker, more complex and harder to constrain.

\subsection{Nebular Spectra and Progenitor Mass}\label{sec:neb_mass}

\begin{figure*}
\begin{minipage}{\textwidth}
\includegraphics[width=180mm]{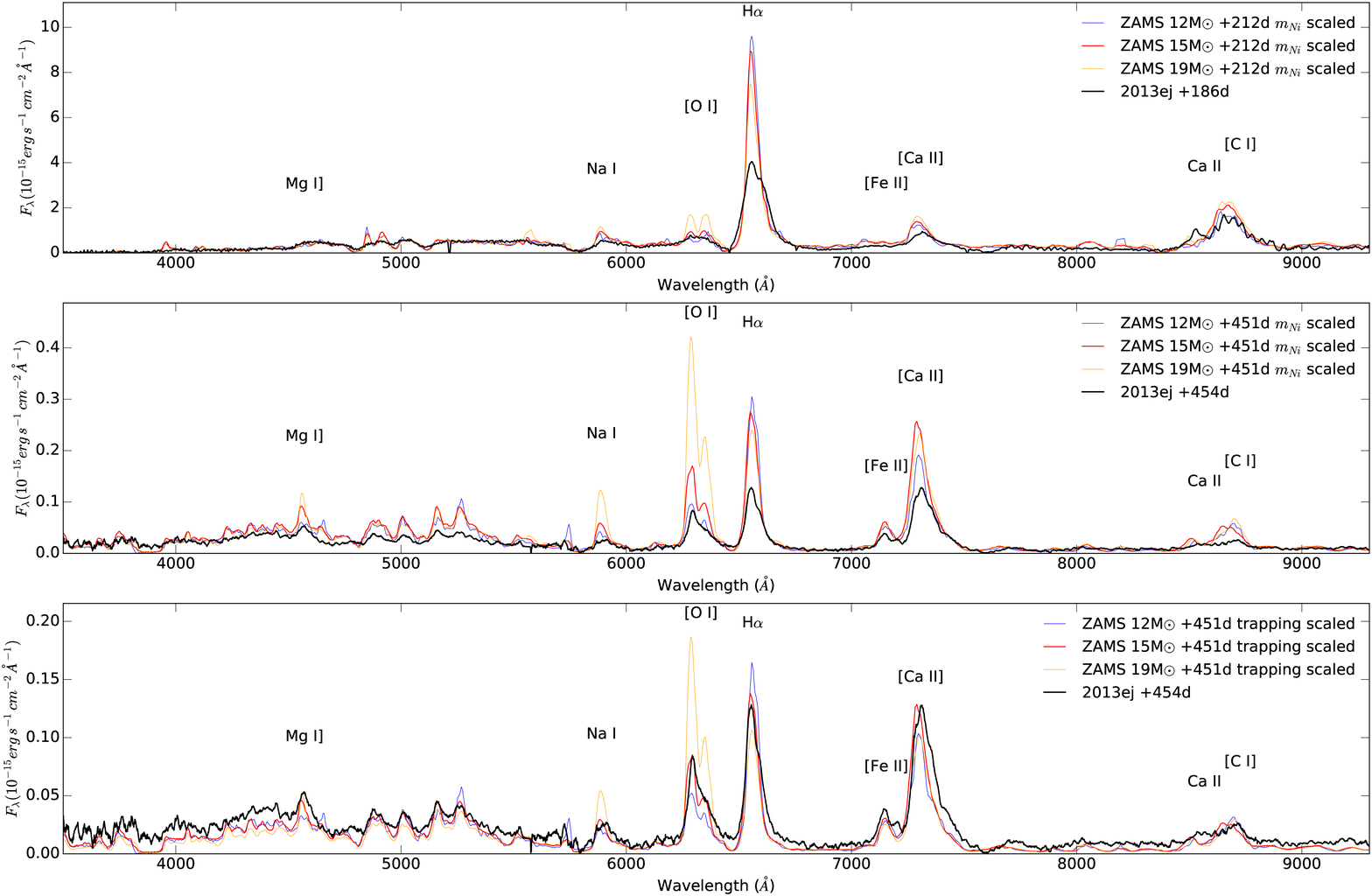}
\caption{Optical spectra of SN 2013ej compared to synthesized models from \citet{2014MNRAS.439.3694J} for 12, 15 and 19 {\msun} (ZAMS) progenitors.
Spectra of SN 2013ej (at 186 and 454 days after explosion) are de-reddened and flux calibrated using multi-band photometry interpolated or extrapolated (linearly) to those epochs. 
In the top and the middle panels, model fluxes are rescaled to the observed epochs using the ratio of {\ni} mass and the model flux decay rate.
In the bottom panel, models fluxes are further rescaled by the relative fraction of trapped $\gamma$-rays.}
\label{fig:neb_opt}
\end{minipage}
\end{figure*}

\citet{2014MNRAS.439.3694J} have shown that a number of nebular lines formed in the core are sensitive to the ZAMS mass of the progenitor. 
We compare our late time optical spectra to models from \citet{2014MNRAS.439.3694J} in Figure~\ref{fig:neb_opt}.
Spectra are flux calibrated using interpolated or (linearly) extrapolated photometry.
The day 186 spectrum is constructed from two spectra taken on day 183 and day 190, calibrated to a common photometry epoch.
Flux calibration is less dependent on interpolation or extrapolation for some other epochs (e.g. day 139 and day 388).
Comparison using those spectra yield the same results.

The model spectra are generated for 0.062 {\msun} of \ni, so we rescale the model by the ratio of {\ni} mass (0.023/0.062), in addition to adjustment for distance.
\citet{2015MNRAS.448.2482J} have shown that change of {\ni} mass (by a factor of two) does not change the spectral shape significantly.
Since the models are not generated on the same days as our observations, we further rescale the fluxes according to the model flux decline rate.
This last adjustment is small for day 186 and negligible for day 454.

The {\oi} $\lambda\lambda$6300, 6364 emission is the most important diagnostic line for the core mass, 
because it is a distinct (unblended) strong line, reemitting a large fraction of the thermalised energy in the O/Ne/Mg zone that contains most of the synthesized oxygen mass.
It is formed close to local thermal equilibrium and insensitive to ionization condition \citep{2012A&A...546A..28J,2014MNRAS.439.3694J}.
In the middle panel of Figure~\ref{fig:neb_opt}, we find that the 12 {\msun} (ZAMS) model matches the strength of {\oi} $\lambda\lambda$6300, 6364 well; 
while the 15 and 19 {\msun} models over-estimate the strength of the line. 
The observed \mgi] $\lambda$4571 and {\naid} lines are weak, also in favour of the lower mass model.

All models over-predict the peak strength of {\ha}, [\caii] $\lambda\lambda$7291, 7323 and the {\caii} NIR triplet, especially at late times.
Our ability to infer the progenitor core mass does not directly depend on these lines as they are mainly formed in the hydrogen-rich zone.
The observed line profiles may be explained by less inward mixing of the hydrogen-rich envelope as a result of a smaller envelope.
Such a smaller envelope is less efficient in trapping $\gamma$-rays and causes a faster flux decline than the lowest mass (12 {\msun}) model.

Alternatively, the smaller trapped $\gamma$-ray fraction can be a result of unusually strong outward mixing of {\ni}.
In such a case, the strength of {\oi} emission may be affected, but the exact effect needs to be carefully modeled.
In the bottom panel of Figure~\ref{fig:neb_opt}, we attempt to rescale the model fluxes further by the observed trapping fraction.
We use a $t_1$ of 313 days for SN 2013ej (as appropriate for the late observations) and a {\ni} mass of 0.020 {\msun} to be consistent with this model.
The effective $t_1$ are 440, 470 and 530 days for our 12, 15 and 19 {\msun} model respectively.
In this case, the 15 {\msun} model provides the best fit overall and for {\oi} $\lambda\lambda$6300, 6364.

Taking into account the uncertainties, models with a progenitor mass between 12 and 15 {\msun} (with 0.3 -- 0.8 {\msun} of oxygen) are favoured. 
Models with progenitor masses much larger than 15 {\msun} are not supported in any case.

\subsection{Emission lines in NIR}\label{sec:nir_lines}

\begin{figure*}
\begin{minipage}{\textwidth}
\includegraphics[width=180mm]{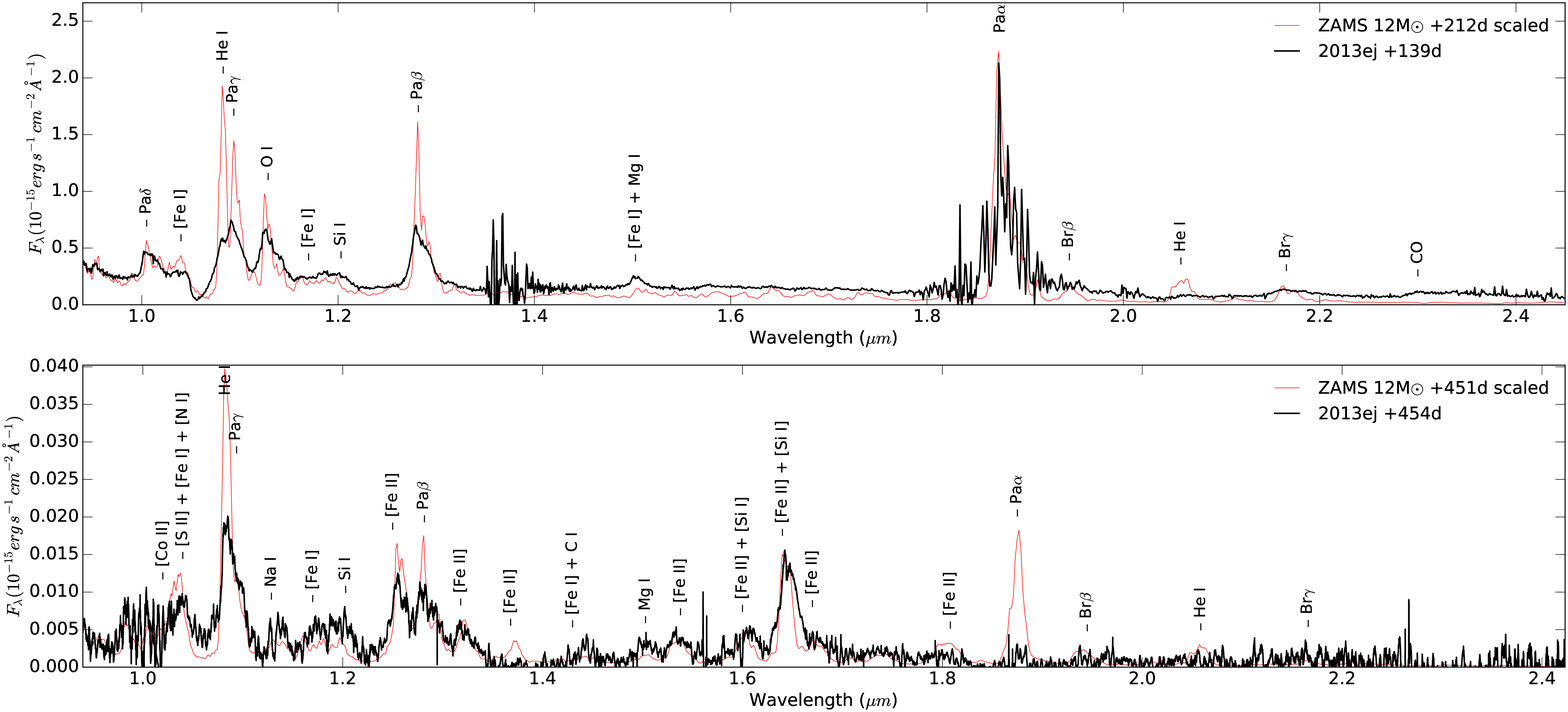}
\caption{NIR spectrum of SN 2013ej at 139 and 454 days after explosion. Spectrum at 139 day is calibrated using $JHKs$ photometry acquired around the same time. 
Spectrum at 454 day is calibrated using the flux scale estimated from the optical part of the X-shooter spectrum.
Spectral features are identified by comparing to the model for a 12{\msun} (ZAMS) progenitor at 212 and 451 days after explosion.
Model fluxes are rescaled to match the observed epochs using the ratio of {\ni} mass and the model flux decay rate.}
\label{fig:neb_nir}
\end{minipage}
\end{figure*}

We have obtained two NIR spectra during the light curve tail at 139 and 454 days after the explosion. 
The spectrum at day 139 is calibrated using $JHKs$ photometry acquired around the same time. 
The spectrum at day 454 is calibrated using a flux scale estimate from the optical part of the X-shooter spectrum.
By comparing our observations to the preferred 12 {\msun} model from \citet{2014MNRAS.439.3694J}, we identify emission features in the NIR wavelength range in Figure~\ref{fig:neb_nir}.

Model fluxes are rescaled in the same way as in the top two panels of $\S$~\ref{sec:neb_mass}, correcting for {\ni} mass and phases, but not the difference in $\gamma$-ray escape rate.
At day 139, the outer ejecta are not yet fully transparent. Higher contribution from the continuum is expected, and weaker lines seen in the day 212 model have not developed. 
At day 454, the NIR flux calibration has relatively large uncertainty, but the overall agreement between model and observation is good.

[\sii] 1.64 $\rm \mu$m is another important diagnostic line for the core mass \citep{2014MNRAS.439.3694J}.
It is blended with [\feii] that is dependent on the {\ni} mass.
The observed strength of the [\feii] + [\sii] 1.64 $\rm \mu$m feature is comparable to the 12 {\msun} or a 15 {\msun} model although the line width is slightly larger than the models.
Good agreement for other [\feii] and [\sii] features suggest the observed {\ni} and Si distribution is roughly consistent with the models.

\subsection{Carbon Monoxide}\label{sec:co}

The first overtone band of CO is clearly detected at 2.3 $\rm \mu$m in the day 139 NIR spectrum. 
At day 454, the spectrum has low signal-to-noise in the $K$-band and shows no sign of any feature at this wavelength.
Formation of molecules is believed to be necessary for dust condensation in the cool ejecta.
Detection of CO molecule is common among Type II SNe \citep[e.g.][]{1988Natur.334..327S,2001A&A...376..188S,2005ApJ...628L.123K,2006ApJ...651L.117K,2006MNRAS.368.1169P,2010MNRAS.404..981M}
and has been reported for Type IIn \citep[SN 1998S;][]{2000AJ....119.2968G, 2001MNRAS.325..907F}, Type IIb \citep[SN 2011dh;][]{2015A&A...580A.142E} and Type Ic SNe \citep{2002PASJ...54..905G,2009A&A...508..371H}.
While the first CO detection varies from SN to SN, it is typically between 100 and 200 days.

\begin{figure}
\includegraphics[width=84.4mm]{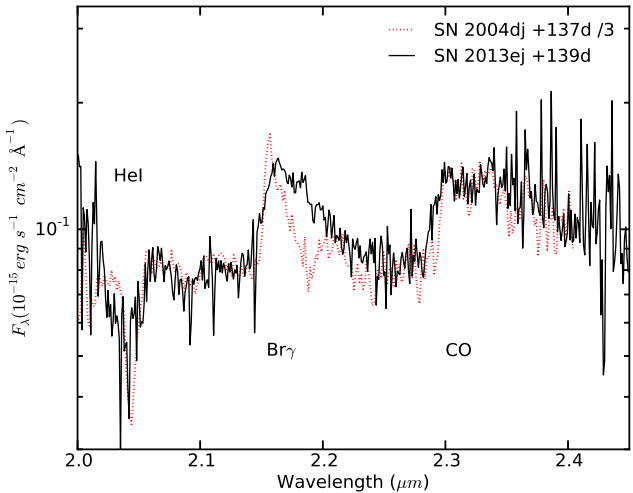}
\caption{Detection of CO first overtone at day 139, compared to observation at a similar epoch for SN 2004dj \citep{2005ApJ...628L.123K}. Observed flux density of SN 2004dj is reduced by a factor of 3 to be comparable to SN 2013ej.}
\label{fig:co}
\end{figure}

In Figure~\ref{fig:co}, we show that the CO feature has a similar shape to that observed for SN 2004dj \citep{2005ApJ...628L.123K}. 
The other prominent feature in this wavelength range is Brackett $\gamma$. Like the other H lines, a significantly broader profile is observed for SN 2013ej than for SN 2004dj.

\section{Discussions}\label{sec:discussion}

In $\S$\ref{sec:bolo}, we have modeled the optical-NIR\footnote{For the period of our concern, contribution from NUV is negligible.} bolometric light curve as powered by {\co} decay with incomplete trapping of $\gamma$-rays.
In $\S$\ref{sec:neb_mass}, we have used synthesized spectral models developed in \citet{2014MNRAS.439.3694J} to explore the progenitor properties.
Here, we further discuss the implications and limitations of these results.

The light curve of SN 2013ej steadily declined during the photospheric phase.
During the nebular phase, the observed flux also declined faster than expected when $\gamma$-rays from {\co} decay are fully trapped.
Relatively fast decays during the tail have been observed for other Type IIL SNe \citep[e.g.][Terreran et al. 2016 in prep]{2014ApJ...786...67A}.
If the decay tail is powered by {\co}, a fast decline rate suggests higher leakage of  $\gamma$-rays, 
resulting from a small ejecta mass, a large kinetic energy, extreme outward mixing of {\ni} or a mixture of these effects.

A small ejecta mass means higher mass loss compared to the single star evolution models we adopt.
Mass loss can be affected by metallicity.
SN 2013ej is close to two {\hh} regions whose emission line abundances have been measured by \citet{2011MNRAS.415.2439R}. 
The mean metallicity of the two regions (and mean of different calibration methods)  is 12+log(O/H) = 8.73.
This is close to the solar value employed by the stellar evolution modeling \citep{2007PhR...442..269W}
for the SN models studied here, although measurements from different empirical calibrations have a root-mean-square spread of 0.2 dex.
If the mass loss is indeed higher, an additional mechanism or a different mass loss prescription is required.
Substantial mass loss can also occur if the star evolved with a companion and it is known that a significant fraction of massive stars are in binary systems \citep[e.g.][]{2012Sci...337..444S}.

Wide nebular emission profiles are observed for SN 2013ej, but velocities are not particularly large during the photospheric phase.
Our model assumes a kinetic energy of $1.2\times10^{51}$~erg and a metal core region expanding with $1800~\rm{km}~\rm{s}^{-1}$,
producing {\oi} emission with comparable width to our observations.

The models assume that {\ni}
is fully macroscopically mixed with the other metal regions in a core
extending to $1800~\rm{km}~\rm{s}^{-1}$. This treatment is based on the limit of strong
mixing seen in multi-dimensional simulations.
To explore the exact impact of a different assumption for mixing, the underlying model needs to be revised which is outside the scope of this paper.

In addition, the models have been developed for a higher {\ni} mass production of 0.062 {\msun}  and scaling down by a factor of larger than two has not been tested. 

Last but not least, our model assumes spherical symmetry which is not supported by the observations.
Significant polarization has been detected a week after the explosion \citep{2013ATel.5275....1L} and also around 100 days \citep{2016MNRAS.456.3157K}, implying asymmetries in both the outer and inner ejecta.
Asymmetric (jet-like) distribution of {\ni} may explain the extra outward mixing and the broad two-component nebular emission lines.

Despite the afore-mentioned caveats, we believe the core mass indicators, such as the {\oi} $\lambda\lambda$6300, 6364 emission, still provide reasonable constraints.
Our spectral models do not reproduce well the observed {\ha} and Ca features. These lines are formed mainly in the hydrogen-rich zone thus more sensitive to choice of mass loss and mixing.
The weaker and flatter-topped {\ha} and Ca emissions are consistent with the result of a weaker reverse shock that causes less mixing of hydrogen-rich material into the core. 

Based on the observed strength of the {\oi} $\lambda\lambda$6300, 6364 feature, 
we find that SN 2013ej has a progenitor with ZAMS mass between 12 and 15 {\msun}.
A progenitor mass much higher than 15 {\msun} produces strong {\oi} and {\nai} that are hard to reconcile with a change of overall scaling.

Our 12 and 15 {\msun} models have ejecta masses of 9.3 and 10.9 {\msun} respectively \citep[see Table~2 in ][]{2012A&A...546A..28J}.
Using their effective decay times ($t_1$) of 440 and 470 days, the simplified Eq~(4) over-estimates their ejecta masses by 20 to 30\%. 
Eq~(4) assumes uniform density and a centralized {\ni} distribution. 
In the model, a dense core increases the optical depth and consequently $t_1$.
This is partly offset by the mixed-out {\ni} but overall Eq~(4) tends to over-estimate mass.
Keeping this in mind, we use Eq~(4) to gauge the meaning of $t_1$.
Fitting the early (< $\sim$170 days) and the late (up to 454 days) decay tail, we obtain two different decay time scales.
The early tail implies a faster decline ($t_1=170$~days) and an ejecta mass of only 5 {\msun} with typical kinetic energy and mixing.
Significant mass loss is required if the progenitor has a ZAMS mass of more than 12 {\msun}.
However, this model underestimates the luminosity measured around 400 days and we find no spectral evidence that the late light curve is affected by other energy sources such as CSM interaction or light echo.  
It is possible that the early tail is affected by residual radiation from the envelope.
Using the late decay tail, we measure a decline rate ($t_1=313$~days) that is still faster than our 12 {\msun} model,
but the discrepancy is not as extreme.

Overall, we find that a progenitor with ZAMS mass between 12 and 15 {\msun} can reasonably produce the observed properties of SN 2013ej.
This mass range is consistent with the estimate from the direct detection \citep{2014MNRAS.439L..56F} and other studies of this SN \citep{2015ApJ...806..160B,2015ApJ...807...59H}.
This progenitor mass also falls in the range of mass estimated for Type IIP SNe from direct detections \citep{2009ARA&A..47...63S} and spectral modeling \citep{2012A&A...546A..28J,2013MNRAS.tmp.1813T,2014MNRAS.439.3694J,2015MNRAS.448.2482J,2016arXiv160308953V}. 
This is further indication that type IIL SNe are unlikely to 
be from progenitors with masses 17--30 {\msun}, and therefore
not likely to explain the lack of high mass progenitor stars 
for nearby core-collapse SNe \citep{2015PASA...32...16S}.

\section{Conclusions}\label{sec:conclusion}

SN 2013ej is classified as a Type IIL based on its relatively fast decline ($\sim$1.7 mag per 100 days in $V$-band) following the initial peak at M$_V$ = -17.6 mag.
Its light curve has a shape that is typical of a Type II SN.
Around $\sim$100 days after the explosion, the luminosity dropped steeply. This feature is a characteristic of Type IIP SNe and has been observed in 
Type IIL SNe that are photometrically monitored for sufficiently long periods \citep{2015MNRAS.448.2608V}.

During the brief transition from a H-recombination dominated phase to the decay tail, we find the $B-V$ colour of SN 2013ej and a few other well-observed Type II SN evolve differently.
The $B$-band covers the wavelength range of many metal lines. 
If this phase traces the photosphere through a transitional layer between the hydrogen-rich envelope and the helium core, the diversity may be due to differences in the line forming regions.

Using the bolometric light curve, we estimate the {\ni} mass to be $0.023\pm0.003$ {\msun}.
After the steep drop-off, the light curve of SN 2013ej declined faster than expected from {\co} decay, indicating a relatively small ejecta mass, a high kinetic energy and/or extended outward mixing of {\ni}.

Our optical and NIR spectral observations both extend to more than a year after the explosion.
We detected the first overtone band of CO at 2.3 $\rm \mu$m at day 139 but the feature is no longer visible at day 454.

We observed broad and asymmetric nebular emission lines.  {\ha} and [\caii] have similarly enhanced red wings.
Asphericity may be a common feature among core-collapse SNe \citep[e.g.][]{2009MNRAS.397..677T} and has played a key role in the SN 2013ej explosion.

We compare our nebular observations to synthesized spectral models \citep{2014MNRAS.439.3694J}.
These models have successfully reproduced the observed spectra of a number of Type IIP SNe.
but have not yet been applied to Type IIL SNe for which late time ($>\sim$ 1 year) observations are sparse (see also Terreran et al. 2016 in prep).
We find that a 12 -- 15 {\msun} (ZAMS) progenitor is preferred for SN 2013ej.
This result is based on comparison of emission lines most sensitive to the core mass of the progenitor thus not prone to uncertainties of mass loss.
This inferred progenitor mass range is similar to the mass estimated for Type IIP SNe from direct detections \citep{2009ARA&A..47...63S} and spectral modeling.

We observed unusually weak
nebular lines formed in the hydrogen envelope (H and Ca lines).
The photometric and spectroscopic behavior of SN 2013ej is consistent
with the idea that Type IIL SNe are formed by explosions of stars
that have lost significant amount of their hydrogen envelope.

\section*{Acknowledgements}
The authors would like to thank Melissa Graham for scheduling the LCOGT observations.
This paper is based on observations collected at the European Organisation for Astronomical Research in the Southern Hemisphere, Chile as part of PESSTO, (the Public ESO Spectroscopic Survey for Transient Objects Survey) ESO program ID 188.D-3003.
The paper is partially based on observations collected at Copernico and Schmidt telescopes (Asiago, Italy) of the INAF - Osservatorio Astronomico di Padova.
Some observations have been obtained also with the 1.22 m telescope + B\&C spectrograph operated in Asiago by the Department of Physics and Astronomy of the University of Padova.
This paper is partly based on observations obtained at the Gemini Observatory, which is operated by the Association of Universities for Research in Astronomy, Inc., under a cooperative agreement with the NSF on behalf of the Gemini partnership: the National Science Foundation (United States), the National Research Council (Canada), CONICYT (Chile), Ministerio de Ciencia, Tecnolog\'{i}a e Innovaci\'{o}n Productiva (Argentina), and Minist\'{e}rio da Ci\^{e}ncia, Tecnologia e Inova\c{c}\~{a}o (Brazil).

This research was made possible through the use of the AAVSO Photometric All-Sky Survey (APASS), funded by the Robert Martin Ayers Sciences Fund.

Parts of this research were conducted by the Australian Research Council Centre of Excellence for All-sky Astrophysics (CAASTRO), through project number CE110001020.
I.R.S. was supported by the ARC Laureate Grant FL0992131.
SB, AP, NER and GT are partially supported by the PRIN-INAF 2014 project ``Transient Universe: unveiling new types of stellar explosions with PESSTO''.
SSchulze acknowledges support from CONICYT-Chile FONDECYT 3140534, Basal-CATA PFB-06/2007, and Project IC120009 ``Millennium Institute of Astrophysics (MAS)'' of Initiative Cient\'{\i}fica Milenio del Ministerio de Econom\'{\i}a, Fomento y Turismo.
This work was partly supported by the European Union FP7 programme through ERC grant number 320360.
KM acknowledges support from the STFC through an Ernest Rutherford Fellowship.
MS acknowledges support from STFC grant ST/L000679/1 and EU/FP7-ERC grant no [615929].
A.G.Y. is supported by the EU/FP7 via ERC grant no. 307260, the Quantum Universe I- CORE Program by the Israeli Committee for Planning and Budgeting and the Israel Science Foundation (ISF); by Minerva and ISF grants; by the Weizmann-UK Òmak- ing connectionsÓ program; and by Kimmel and ARCHES awards.








\clearpage

\begin{footnotesize}

\tablefirsthead{\hline
\multirow{2}{*}{UT Date} & JD & Magnitude & \multirow{2}{*}{filter} & Telescope/  \\
&( -2400000) & (error) & & Instrument\textsuperscript{a}\\ \hline}

\tablehead{\multicolumn{5}{c}%
{{\bfseries \tablename\ \thetable{} -- continued from previous page}} \\ \hline 
\multirow{2}{*}{UT Date} & JD & Magnitude & \multirow{2}{*}{filter} & Telescope/  \\
&( -2400000) & (error) & & Instrument\textsuperscript{a}\\ \hline}

\tabletail{\hline \multicolumn{5}{c}{{Continued on next page}} \\ \hline}

\tablelasttail{\hline
\multicolumn{5}{p{\linewidth}}{\textsuperscript{a}
\footnotesize{
Optical telescopes and instruments:
1m008: LCOGT 1m at McDonald observatory, USA;
1m010, 1m012, 1m013: LCOGT 1m at Sutherland, South Africa;
1m004, 1m005, 1m009: LCOGT 1m at Cerro Tololo, Cile;
1m003, 1m011: LCOGT 1m at Siding Spring, Australia;
LOT: 1m at the Lulin Observatory in Taiwan;
AFOSC:  INAF 1.82 m + AFOSC;
SBIG: INAF Schmidt 67/92 cm + SBIG;
NTT: NTT + EFOSC2;
VLT: X-shooter Acquisition and Guiding camera.
See $\S$\ref{sec:obs} for further details.}} \\
\hline}

\topcaption{Optical and NIR photometric observations of SN 2013ej.}\label{tab:phot}

\begin{supertabular}{ccccc} 
2013/07/27 & 56500.878 &         12.25 (0.06) &     U &      1m004 \\
2013/07/28 & 56501.878 &         12.16 (0.05) &     U &      1m004 \\
2013/08/03 & 56507.859 &         12.12 (0.06) &     U &      1m008 \\
2013/08/04 & 56508.862 &         12.13 (0.08) &     U &      1m008 \\
2013/08/05 & 56509.932 &         12.21 (0.06) &     U &      1m008 \\
2013/08/06 & 56510.867 &         12.32 (0.05) &     U &      1m008 \\
2013/08/09 & 56513.839 &         12.53 (0.05) &     U &      1m008 \\
2013/08/09 & 56513.876 &         12.51 (0.06) &     U &      1m004 \\
\end{supertabular}
\end{footnotesize}







\bsp	
\label{lastpage}
\end{document}